\documentclass[aps,pre,amsmath,singlecolumn,showpacs,11pt]{revtex4-2}
\usepackage{graphicx}
\usepackage{amssymb,amsmath}
\usepackage{bm}
\usepackage{color}

\usepackage{xcolor} 

\begin{document}

\title{Dimension reduction of dynamical systems on networks with leading and non-leading eigenvectors of adjacency matrices}
\date{\today}
\author{Naoki Masuda$^{1,2}$}
\email{naokimas@buffalo.edu}
\author{Prosenjit Kundu$^1$} 

\affiliation{$^1$Department of Mathematics, State University of New York at Buffalo, NY 14260-2900, USA }
\affiliation{$^2$Computational and Data-Enabled Science and Engineering Program, State University of New York at Buffalo, Buffalo, NY 14260-5030, USA}
\date{\today}

\begin{abstract}
Dimension reduction techniques for dynamical systems on networks are considered to promote our understanding of the original high-dimensional dynamics. One strategy of dimension reduction is to derive a low-dimensional dynamical system whose behavior approximates the observables of the original dynamical system that are weighted linear summations of the state variables at the different nodes. Recently proposed methods use the leading eigenvector of the adjacency matrix of the network as the mixture weights to obtain such observables. In the present study, we explore performances of this type of one-dimensional reductions of dynamical systems on networks when we use non-leading eigenvectors of the adjacency matrix as the mixture weights. Our theory predicts that non-leading eigenvectors can be more efficient than the leading eigenvector and enables us to select the eigenvector minimizing the error. We numerically verify that the optimal non-leading eigenvector outperforms the leading eigenvector for some dynamical systems and networks. We also argue that, despite our theory, it is practically better to use the leading eigenvector as the mixture weights to avoid misplacing the bifurcation point too distantly and to be resistant against dynamical noise.
\end{abstract}

\maketitle

\section{Introduction}

A variety of complex systems in the real world can be described by dynamical systems on networks \cite{Boccaletti2006PhysRep,Barrat2008book,Porter2016book,Newman2018book}. This seems to be the case in particular when systems of question are composed of dynamical elements that are similar to each other except for the connectivity and some easily parameterizable heterogeneity across the individual elements. Examples include coupled oscillators on networks \cite{Arenas2008PhysRep} including models of functioning of power grids \cite{Motter2013NatPhys}, predator-prey, mutualistic, and other dynamics impacting community stability in ecological networks \cite{Thebault2010Science,Allesina2012Nature}, gene regulatory networks \cite{Alon2007book}, and epidemic processes and ecological population dynamics considered on networks of habitat patches (called metapopulation models) \cite{Hanski1998Nature,Colizza2007NatPhys}. Analyses of dynamical systems on networks have clarified various collective phenomena on networks such as phase transitions and synchronous oscillations.

Dynamical systems on networks are necessarily high-dimensional because each node is assigned with one or more dynamical variables and those variables interact via edges of the given network. Even if the dynamics at each node is one-dimensional, the entire dynamical system on the network is $N$-dimensional, where $N$ is the number of nodes in the network.
Therefore, similar to various dimension reduction techniques for high-dimensional data \cite{Hinton2006Science,Bengio2013IeeeTransPatAnalMachineIntel,Cunningham2015JMachineLearnResearch}, one may be tempted to
map a dynamical system on networks into one in a low-dimensional space without losing much information. Then, by deploying analytical and numerical techniques suited to low-dimensional dynamical systems, we may be better able to understand the original high-dimensional network dynamics.
Gao, Barzel, and Barab\'{a}si developed a heterogeneous mean-field theory, assuming uncorrelated networks with general degree (i.e., number of edges that a node in the given network has) distributions, to reduce a class of dynamical systems on networks to one-dimensional dynamical systems \cite{GaoBarzelBarabasi2016Nature}. Their method approximates a one-dimensional projection of the original high-dimensional dynamics on networks with a reasonably good accuracy in various cases.
See Refs.~\cite{Tu2017PhysRevE,KunduKoriMasuda2022PhysRevE} for validation studies of this approach and Ref.~\cite{Tu_IScience2021} for an extension.
Furthermore, to deal with general network structure, Laurence \textit{et al.} developed a systematic method, which we call the spectral method, to use the eigenvalues and eigenvectors of the adjacency matrix of the network to reduce the same class of dynamics on networks into dynamics of small dimensions, such as one or two \cite{Laurence2019PhysRevX}. See Ref.~\cite{Thibeault2020PhysRevResearch} for a further advancement of the spectral method. In the case of one-dimensional reduction, which we focus on in the present study, the spectral method uses an observable that is a particular
linear combination of the state variables on all nodes, denoted by $R = \sum_{i=1}^N a_i x_i$, where $x_i$ is the dynamical state of the $i$th node, and $a_i$ is the mixing weight. Then, one writes down a closed dynamical equation in terms of $R$.
On a theoretical basis, they proposed to use the $i$th element of the leading eigenvector (i.e., the eigenvector associated with the largest eigenvalue) of the adjacency matrix as $a_i$
\cite{Laurence2019PhysRevX}. 

The leading eigenvector of the adjacency matrix is a key descriptor of contagious processes on networks because the adjacency matrix tells us who can directly infect whom. In fact, the $i$th element of the leading eigenvector gives the likelihood that the $i$th node is infectious when the infection rate of an epidemic process model such as the susceptible-infectious-susceptible (SIS) model is poised near the epidemic threshold~\cite{Goltsev2012PhysRevLett}. For example, scale-free networks (i.e., networks with power-law degree distributions) show eigenvector localization such that the leading eigenvector has only a small fraction of considerably positive elements, which are at the largest-degree nodes and correspond to the presence of infection at these nodes~\cite{Goltsev2012PhysRevLett,Martin2014PhysRevE,Pastorsatorras2016SciRep,Pastorsatorras2018JStatPhys}. In contrast, the other elements of the eigenvector are close to 0, corresponding to the scarcity of infection at small-degree nodes. These and other results lend support for the spectral method \cite{Laurence2019PhysRevX} and its extension called the dynamics approximate reduction technique (DART) \cite{Thibeault2020PhysRevResearch}, which use the leading eigenvectors of the adjacency matrix as mixing weights.

In the present study, we develop a theory to argue that it is often better to use a non-leading eigenvector of the adjacency matrix for the spectral method to realize a better accuracy at reducing the original $N$-dimensional dynamics into a one-dimensional dynamics. We derive optimization criteria under the assumption that $\{x_1, \ldots, x_N\}$ is not too heterogeneous, which numerically holds true for various dynamical systems on networks \cite{KunduKoriMasuda2022PhysRevE}. Our theoretical derivation suggests that the leading eigenvector does not necessarily yield the smallest error of the dimension reduction by the spectral method. For various networks, a non-leading eigenvector, which implies that we linearly combine $x_i$ into one observable with some negative weights $a_i$, realizes a smaller error than the spectral method with the leading eigenvector. We verify our theory by numerical simulations of three dynamical systems on different networks. Finally, we argue that, despite our theory, the spectral method using the leading eigenvector as the mixing weights is better than with the non-leading eigenvector because of two factors that our theory does not address: precision in locating the bifurcation point and the robustness against dynamical noise.
Our code for computing the optimal eigenvectors and reproducing the results in this article is available at
\url{https://github.com/naokimas/nonleading-spectral}.

\section{Spectral method}

Throughout the present study, we consider the following class of dynamical systems on networks \cite{Barzel2013NatPhys,GaoBarzelBarabasi2016Nature,Laurence2019PhysRevX}:
\begin{equation}
\frac{\text{d}x_i}{\text{d}t} = F(x_i) + \sum_{j=1}^N w_{ij} G(x_i, x_j),
\label{eq:system}
\end{equation}
where $t$ is the time, $x_i$ is the one-dimensional dynamical state of the $i$th node (with $i\in \{1, \ldots, N\})$, $F(x)$ represents the intrinsic dynamics of the node, $G(x_i, x_j)$ represents the influence of $x_j$ on $x_i$, and $w_{ij}$ is the strength of the influence of node $j$ on node $i$, corresponding to the weighted adjacency matrix of the given network. We assume that the network is connected. We also assume that the network does not have self-loops, i.e., $w_{ii} = 0$ for $i\in \{1, \ldots, N\}$.
However, if all nodes have a self-loop of the same edge weight (i.e., $w_{11} = \cdots = w_{NN}$), one can include the effect of such self-loops into $F(x_i)$ by replacing the original $F(x_i)$ by $F(x_i) + w_{ii} G(x_i, x_i)$.

We describe the spectral method \cite{Laurence2019PhysRevX} in this section. With this method, one reduces the $N$-dimensional dynamical system given by Eq.~\eqref{eq:system}
to an $n$-dimensional system, where $n\ll N$, by deriving an approximate $n$-dimensional dynamical system in terms of observables each of which is a linear combination of $\{ x_1, \ldots, x_N \}$. We focus on the case of $n=1$ in this paper. We consider an observable, which we denote by $R$, given by
\begin{equation}
R = \sum_{i=1}^N a_i x_i,
\label{eq:R-def}
\end{equation}
where $\{a_1, \ldots, a_N \}$ is normalized such that $\sum_{i=1}^N a_i = 1$.

By combining Eqs.~\eqref{eq:system} and \eqref{eq:R-def}, one obtains
\begin{equation}
\frac{\text{d}R}{\text{d}t} = \sum_{i=1}^N a_i \left[ F(x_i) + \sum_{j=1}^N w_{ij} G(x_i, x_j) \right].
\label{eq:dR/dt}
\end{equation}

By Taylor expanding $F(x_i)$ around $x_i = R$ to the first order, we obtain
\begin{equation}
F(x_i) = F(R) + (x_i-R) F'(R) + O\left((x_i - R)^2\right).
\label{eq:F-Taylor}
\end{equation}
Similarly, we expand $G(x_i, x_j)$ around $x_i = \beta R$ and $x_j = \gamma R$, where $\beta$ and $\gamma$ are constants to be determined, to obtain
\begin{equation}
G(x_i, x_j) = G(\beta R, \gamma R) + (x_i - \beta R) G_1 (\beta R, \gamma R) + (x_j - \gamma R) G_2(\beta R, \gamma R) + O\left((x_i - \beta R)^2\right) + O\left((x_j - \gamma R)^2\right),
\label{eq:G-Taylor}
\end{equation}
where $G_1$ and $G_2$ are the partial derivatives of $G$ with respect to the first and second argument, respectively.
By substituting Eqs.~\eqref{eq:F-Taylor} and \eqref{eq:G-Taylor} into Eq.~\eqref{eq:dR/dt}, one obtains
\begin{align}
\frac{\text{d}R}{\text{d}t} =& F(R) + \alpha G(\beta R, \gamma R) + G_1(\beta R, \gamma R) \sum_{i, j=1}^N a_i w_{ij} (x_i - \beta R)
+ G_2(\beta R, \gamma R) \sum_{i, j=1}^N a_i w_{ij} (x_j - \gamma R)\notag\\
& + O\left((x - R)^2\right) + O\left((x - \beta R)^2\right) + O\left((x - \gamma R)^2\right),
\label{eq:dR/dt-with-G_1-and-G_2}
\end{align}
where
\begin{equation}
\alpha = \sum_{i, j=1}^N a_i w_{ij},
\label{eq:def-alpha}
\end{equation}
and $O\left((x - R)^2\right)$ is a short-hand notation for $\sum_{i=1}^N O\left((x_i - R)^2\right)$ and similar for 
$O\left((x - \beta R)^2\right)$ and $O\left((x - \gamma R)^2\right)$. 
By requiring that the first-order terms in Eq.~\eqref{eq:dR/dt-with-G_1-and-G_2}, i.e., those containing $G_1$ and $G_2$, disappear for any $\{ x_1, \ldots, x_N \}$, one obtains
\begin{align}
\alpha \beta R =& \sum_{i, j=1}^N a_i w_{ij} x_i = \bm x^{\top} K \bm a,
\label{eq:condition-1}\\
\alpha \gamma R =& \sum_{i, j=1}^N a_i w_{ij} x_j = \bm x^{\top} W^{\top} \bm a,
\label{eq:condition-2}
\end{align}
where $\bm x = (x_1, \ldots, x_N)^{\top}$, $\bm a = (a_1, \ldots, a_N)^{\top}$, $K$ is the $N\times N$ diagonal matrix whose $i$th diagonal entry is equal to the weighted in-degree of the $i$th node, i.e., $k_i^{\text{in}} \equiv \sum_{j=1}^N w_{ij}$, $W = (w_{ij})$ is the $N\times N$ adjacency matrix, and ${}^{\top}$ represents the transposition. By substituting Eq.~\eqref{eq:R-def} in Eqs.~\eqref{eq:condition-1} and \eqref{eq:condition-2}, one obtains
\begin{align}
\bm x^{\top} (\alpha \beta \bm a - K \bm a) =& 0,
\label{eq:condition-1-p}\\
\bm x^{\top} (\alpha \gamma \bm a - W^{\top} \bm a) =& 0.
\label{eq:condition-2-p}
\end{align}
Because Eqs.~\eqref{eq:condition-1-p} and \eqref{eq:condition-2-p} ideally hold true for any $\bm x$, one obtains
\begin{align}
K \bm a =& \alpha \beta \bm a,
\label{eq:condition-1-pp}\\
W^{\top} \bm a =& \alpha \gamma \bm a.
\label{eq:condition-2-pp}
\end{align}
Equations~\eqref{eq:condition-1-pp} and \eqref{eq:condition-2-pp} indicate that $\bm a$ is a right eigenvector of both $K$ and $W^{\top}$. However,
$K$ and $W^{\top}$ do not share the eigenspace in general. In particular, the eigenvectors of $K$ are the standard unit vectors because $K$ is a diagonal matrix. The standard unit vector $\bm a = (1, 0, \ldots, 0)^{\top}$, for example, satisfies
Eq.~\eqref{eq:condition-1-pp}, but it implies that $R = x_1$ so that we only observe the first node. Note that $\bm a = (1, 0, \ldots, 0)^{\top}$
satisfies Eq.~\eqref{eq:condition-2-pp} if and only if $k_1^{\text{in}} = 0$ such that the first node is not influenced by any other node.

Laurence \textit{et al.}~proposed to set $\bm a$ to be a right eigenvector of $W^{\top}$. Because $\bm a$ is normalized such that $\bm 1^{\top} \bm a = 1$, where $\bm 1 = (1, \ldots, 1)^{\top}$, we left-multiply $\bm 1^{\top}$ by Eq.~\eqref{eq:condition-2-pp} and use $\bm 1^{\top} W^{\top} = \bm k^{\rm in}$, where 
$\bm k^{\rm in} = (k_1^{\rm in}, \ldots, k_N^{\rm in})^{\top}$, and $\alpha = \bm a^{\top} \bm k^{\rm in}$, which originates from Eq.~\eqref{eq:def-alpha}, to obtain $\alpha = \alpha \gamma$, i.e., $\gamma = 1$.
Substitution of $\gamma=1$ in Eq.~\eqref{eq:condition-2-pp} implies that $\alpha$ is the eigenvalue of $W^{\top}$ associated with eigenvector $\bm a$.
One can normalize $\bm a$ as $\bm 1^{\top} \bm a = 1$ unless the associated eigenvalue is 0.
Finally, because Eq.~\eqref{eq:condition-1-pp} no longer holds true in general, one uses the $\beta$ value that minimizes the approximation error, i.e.,
\begin{equation}
\beta^* = \text{argmin}_{\beta} \left\| K \bm a - \alpha \beta \bm a \right\|^2 = \frac{\bm b^{\top} \bm k^{\rm in}} {\bm a^{\top} \bm k^{\rm in}},
\label{eq:beta^*-Laurence}
\end{equation}
where $\bm b = (b_1, \ldots, b_N)^{\top}$, and $b_i = a_i^2 / \sum_{\ell=1}^N a_{\ell}^2$ with $i\in \{1, \ldots, N\}$.

By neglecting the second and higher order terms in 
Eq.~\eqref{eq:dR/dt-with-G_1-and-G_2}, one obtains the spectral reduction given by
\begin{equation}
\frac{\text{d}R}{\text{d}t} = F(R) + \alpha G(\beta^* R, R).
\label{eq:Laurence-final}
\end{equation}

\section{Modified spectral method}

\subsection{Eigenvector minimizing the approximation error\label{sub:our-theory}}

\begin{figure}
\includegraphics[width=0.475\textwidth]{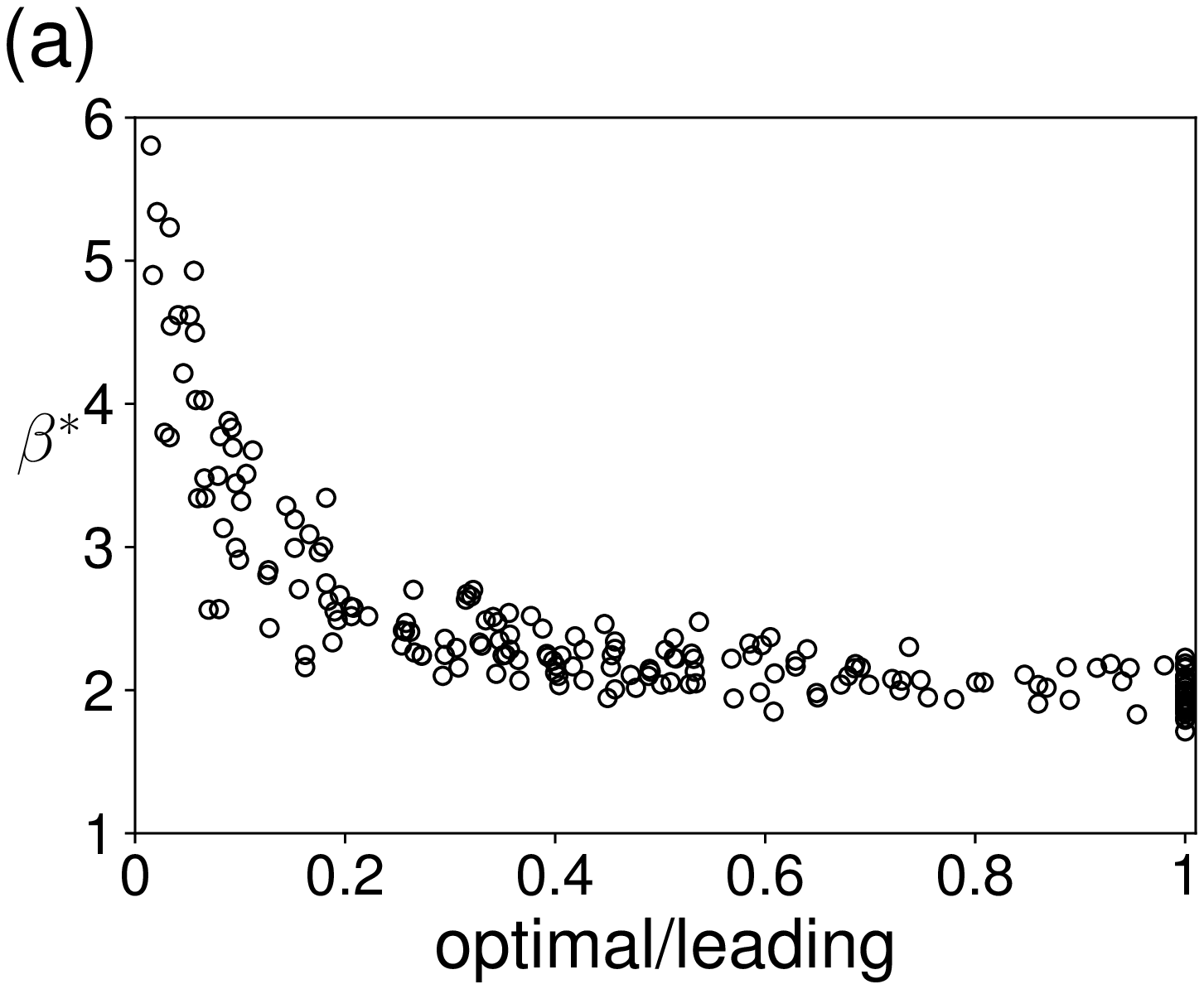}
\includegraphics[width=0.475\textwidth]{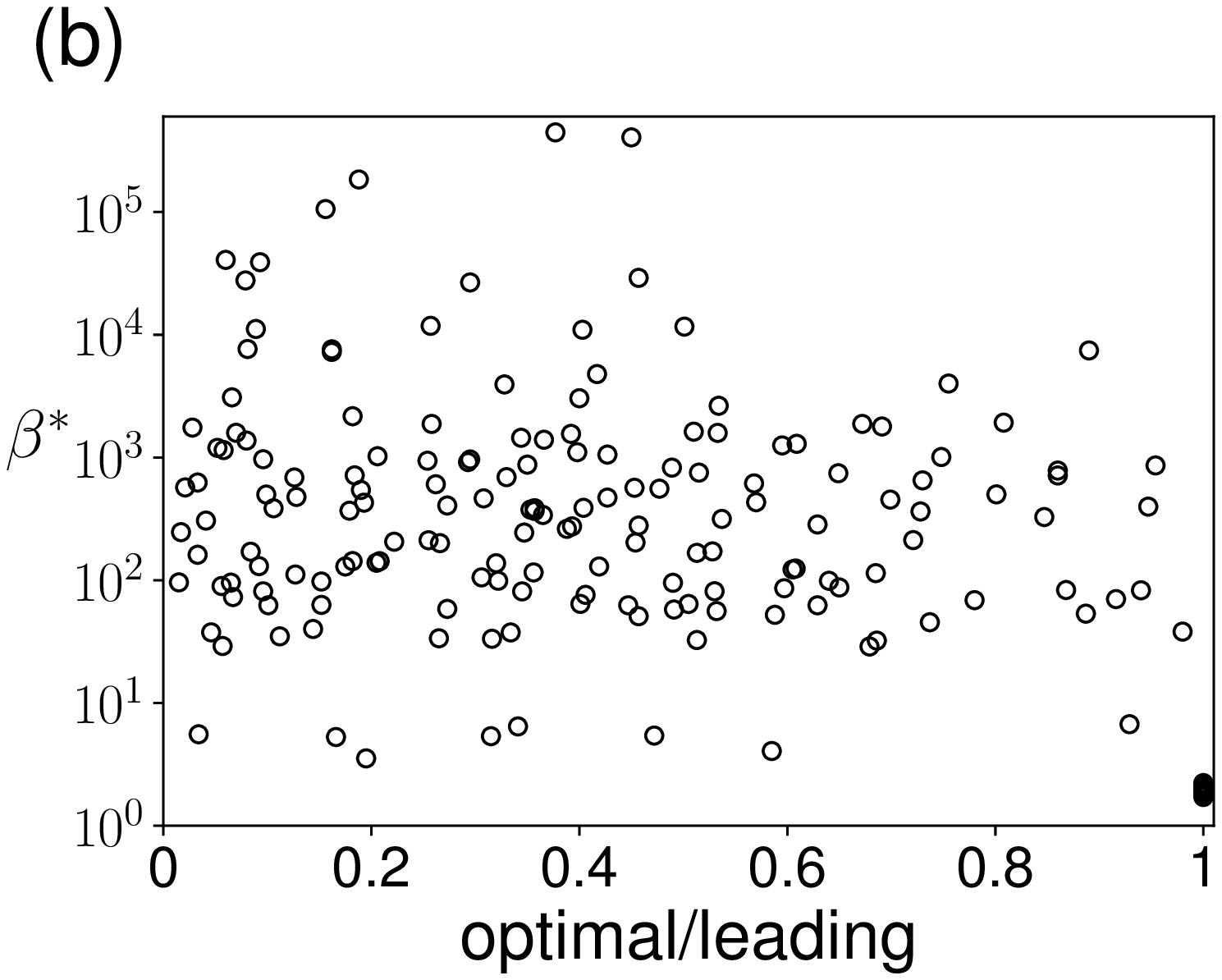}
\caption{The $\beta^*$ value for $200$ scale-free networks (i.e., networks with power-law degree distributions) with $N=1000$ nodes.
Each circle represents a network.
The horizontal axis represents the $\epsilon_1$ value for the minimizer of $\epsilon_1$ divided by the $\epsilon_1$ value for the leading eigenvector. A small value on the horizontal axis implies that the minimizer of $\epsilon_1$ is efficient at reducing $\epsilon_1$ relative to the leading eigenvector. The value on the horizontal axis being equal to 1 implies that the leading eigenvector minimizes $\epsilon_1$.
(a) $\beta^*$ for the leading eigenvector. (b) $\beta^*$ for the eigenvector minimizing $\epsilon_1$.
We use the configuration model to generate the $200$ scale-free networks with the power-law exponent of the degree distribution equal to $\tilde{\gamma} = 3.5$. See Section~\ref{sub:performance-various-networks} for details of the degree distribution used.
Note that one can rewrite Eq.~\eqref{eq:beta^*-Laurence} as 
$\beta^* = \bm b^{\top} \bm k^{\rm in} / \alpha$. Therefore, if the minimizer of $\epsilon_1$ is associated with a small eigenvalue, $\alpha$,
then $\beta^*$ tends to be large. This explains why the $\beta^*$ value tends to be much larger in (b) than (a).
}
\label{fig:beta^*}
\end{figure}

With the optimal $\beta$ value given by Eq.~\eqref{eq:beta^*-Laurence}, we obtain
\begin{equation}
e_1 \equiv \left\| K \bm a - \alpha \beta^* \bm a \right\|^2 = 
\frac{\sum_{i=1}^N \sum_{j=1}^{i-1} a_i^2 a_j^2 (k_i^{\rm in} - k_j^{\rm in})^2}
{\sum_{\ell=1}^N a_\ell^2}.
\label{eq:error-1}
\end{equation}
We have two remarks on Eq.~\eqref{eq:error-1}.
First, although the use of the leading eigenvector as $\bm a$ has been recommended for the spectral method \cite{Laurence2019PhysRevX} and the DART \cite{Thibeault2020PhysRevResearch}, $e_1$ may be smaller for other right eigenvectors of $W^{\top}$.
Second, Eq.~\eqref{eq:error-1} implies that $e_1 = 0$ if all the nodes have the same in-degree, regardless of which right eigenvector of $W^{\top}$ we use as $\bm a$.
However, numerical simulations indicate that the spectral method is not exact even for random regular graphs \cite{KunduKoriMasuda2022PhysRevE}. 
The approximation error may be due to the higher-order terms in the Taylor expansion in Eq.~\eqref{eq:dR/dt-with-G_1-and-G_2} or 
the fact that we have expanded Eq.~\eqref{eq:dR/dt} around both $x_i = R$ and $x_i = \beta^* R$. 
(Although we also expanded $x_i$ around $\gamma R$, we found that $\gamma = 1$.)
In fact, our numerical test suggests that $\beta^*$ is often far from $1$ both for the leading eigenvector and the eigenvector minimizing $\epsilon_1$, as we show for scale-free networks in Fig.~\ref{fig:beta^*}(a) and Fig.~\ref{fig:beta^*}(b), respectively.

For these reasons, we propose to expand Eq.~\eqref{eq:dR/dt} only around $x_i = R$ as follows. Let us set $x_i = R + \Delta x_i$ (with $i\in \{1, \ldots, N\}$), where $\left| \Delta x_i \right| \ll 1$. Then, we obtain
\begin{align}
& \sum_{i, j=1}^N a_i w_{ij} G(x_i, x_j)\notag\\
=& \sum_{i, j=1}^N a_i w_{ij} G\left( R+\Delta x_i, R+\Delta x_j \right)\notag\\
=& \sum_{i, j=1}^N a_i w_{ij} \left[ G(R, R) + G_1(R, R) \Delta x_i + G_2(R, R) \Delta x_j \right] + O\left( (\Delta x)^2 \right)\notag\\
=& \alpha G(R, R) + G_1(R, R) \sum_{i, j=1}^N a_i w_{ij} (x_i - R) + G_2(R, R) \sum_{i, j=1}^N a_i w_{ij} (x_j - R) + O\left( (\Delta x)^2\right)\notag\\
=& \alpha G(R, R) + G_1(R, R) \left( \sum_{i, j=1}^N a_i w_{ij} x_i - \alpha R \right) 
+ G_2(R, R) \left( \sum_{i, j=1}^N a_i w_{ij} x_j - \alpha R \right) + O\left((\Delta x)^2\right).
\label{eq:new-approximation-1}
\end{align}
By imposing that the first-order terms on the right-hand side of Eq.~\eqref{eq:new-approximation-1} disappear, we obtain
\begin{align}
\alpha R =& \bm x^{\top} K \bm a,
\label{eq:new-condition-1}\\
\alpha R =& \bm x^{\top} W^{\top} \bm a.
\label{eq:new-condition-2}
\end{align}
By substituting $R = \bm x^{\top} \bm a$ in
Eqs.~\eqref{eq:new-condition-1} and \eqref{eq:new-condition-2} and imposing that Eqs.~\eqref{eq:new-condition-1} and \eqref{eq:new-condition-2} hold true for any $\bm x$, we obtain
\begin{align}
K \bm a =& \alpha \bm a,
\label{eq:new-condition-1-pp}\\
W^{\top} \bm a =& \alpha \bm a.
\label{eq:new-condition-2-pp}
\end{align}
For the same reason as that for the original spectral method, Eqs.~\eqref{eq:new-condition-1-pp} and \eqref{eq:new-condition-2-pp} do not simultaneously hold true in general. Therefore, by following the recommendation for DART \cite{Thibeault2020PhysRevResearch},
we require Eq.~\eqref{eq:new-condition-2-pp} to hold exactly and let Eq.~\eqref{eq:new-condition-1-pp} be satisfied only approximately.
The opposite case, i.e., to impose Eq.~\eqref{eq:new-condition-1-pp} strictly and Eq.~\eqref{eq:new-condition-2-pp} only approximately, implies that we 
only observe a single node as a representative of the entire system (see Appendix~\ref{sec:observe-one-variable} for the derivation).
Equation~\eqref{eq:new-condition-2-pp} implies that, as in the case of the original spectral method, $\bm a$ must be a right eigenvector of $W^{\top}$ and that $\alpha$ is the associated eigenvalue. Therefore, we select the eigenvalue $\alpha$ and the associated eigenvector $\bm a$ that minimize the
the approximation error for Eq.~\eqref{eq:new-condition-1-pp} defined by
\begin{equation}
\epsilon_2 \equiv \left\| K \bm a - \alpha \bm a \right\|^2 =  \sum_{i=1}^N (k_i^{\rm in} - \alpha)^2 a_i^2.
\label{eq:error-modified-reduction}
\end{equation}
We remind that $\bm a$ is normalized such that $\sum_{i=1}^N a_i = 1$.
The modified one-dimension reduction reads
\begin{equation}
\frac{\text{d}R}{\text{d}t} = F(R) + \alpha G(R, R).
\label{eq:modified-reduction-final}
\end{equation}

A few remarks are in order. First, as in the case of the original spectral method, the leading eigenvector of $W^{\top}$ may not minimize $e_2$.
Second, $\epsilon_2 \ge \epsilon_1$ holds true because our one-dimensional reduction scheme corresponds to $\beta=1$, whereas the original spectral method also optimizes the $\beta$ value. Nevertheless, which reduction method is more accurate than the other is a nontrivial question because the original spectral method uses the Taylor expansion of each $x_i$ around two reference values (i.e., $R$ and $\beta^* R$). Third, the Perron-Frobenius theorem guarantees that $a_i > 0$, $\forall i \in \{1, \ldots, N\}$ when $\bm a$ is the right leading eigenvector of $W^{\top}$. Therefore, the observable, $R$, when the spectral method uses the leading eigenvector, is a weighted average of all the dynamical variables, $x_i$, with positive weights. The Perron-Frobenius theorem also guarantees that all the entries of the left leading eigenvector of $W^{\top}$, which we denote by $\bm v \in \mathbb{R}^{1\times N}$, are also positive. If $W^{\top}$ is diagonalizable and the leading eigenvalue is not repeated, we obtain $\bm v \bm a = 0$ for any right eigenvector $\bm a$ of $W^{\top}$ other than the leading one. Therefore, the sign of at least one $a_i$ must be opposite to the sign of some other $a_i$. This fact implies that $R = \bm a^{\top} \bm x$ is a weighted average of $\{ x_1, \ldots, x_N \}$ including negative weights.

\subsection{Case of regular graphs}

In the case of the complete graph, $K$ is the diagonal matrix whose all diagonal entries are equal to $N-1$. Matrix $W$ is given by
$W=(w_{ij})$, where $w_{ij} = 1 - \delta_{ij}$, and $\delta_{ij}$ is the Kronecker delta. In this case, $W$ has the leading eigenvalue $\alpha = N-1$ with the associated eigenvector $\bm a = (1, \ldots, 1)^{\top}$ and the ($N-1$)-fold eigenvalue $\alpha = -1$ whose associated eigenvectors can be chosen as
$\bm a = (1, -1, 0, \ldots, 0)^{\top}$, $(1, 0, -1, 0, \ldots, 0)^{\top}$, $(1, 0, 0, -1, 0, \ldots, 0)^{\top}$, $\ldots$, $(1, 0, \ldots, 0, -1)^{\top}$. We find that the combination of $\alpha = N-1$ and $\bm a = (1, \ldots, 1)^{\top}$ satisfies Eq.~\eqref{eq:new-condition-1-pp}, whereas the combination of $\alpha = -1$ and any of its eigenvector does not. Therefore, only the leading eigenvalue and eigenvector realize $\epsilon_2 = 0$. Furthermore, due to the symmetry, all the $x_i$ values are same, such that the
$O\left((\Delta x)^2\right)$ in Eq.~\eqref{eq:new-approximation-1} disappears. Therefore, the spectral method is exact. Because the complete graph is a regular graph (i.e., a network in which all the nodes have the same degree), we also obtain $\epsilon_1 = 0$ for any eigenvector.

Similar results hold true for regular graphs in general. Specifically, for any undirected regular graphs with degree $k$, the leading eigenvalue of $W$ is equal to $k$, and the associated eigenvector is $\bm a = (1, \ldots, 1)^{\top}$.
We find that this pair of eigenvalue and eigenvector, which is known to be of multiplicity 1 \cite{Biggs1993book},
%
%
satisfies Eq.~\eqref{eq:new-condition-1-pp}. No other pair of eigenvalue and the associated eigenvector of $W$ does not satisfy Eq.~\eqref{eq:new-condition-1-pp} because Eq.~\eqref{eq:new-condition-1-pp} implies that the eigenvalue needs to be equal to $k$. With the leading eigenvalue and eigenvector of $W^{\top}$, if the network is vertex-transitive \cite{Biggs1993book} (e.g., square lattice with periodic boundary conditions), 
$O\left((\Delta x)^2\right)$ in Eq.~\eqref{eq:new-approximation-1} disappears due to the symmetry (i.e., $x_i = x_j$ for all $i, j \in \{1, \ldots, N\}$) such that the spectral method is exact. Otherwise, the $O\left((\Delta x)^2\right)$ term may cause the discrepancy between the spectral method and the numerical results, as is the case for random regular graphs \cite{KunduKoriMasuda2022PhysRevE}. 

\subsection{Case of degree-heterogeneous random graphs}

Consider undirected configuration models, i.e., uniformly random undirected networks with a given degree sequence. In the limit of $N\to\infty$,
the normalized leading eigenvector of the adjacency matrix is approximated by $a_i = k_i / \sum_{\ell=1}^N k_{\ell}$ \cite{Pastorsatorras2016SciRep}, and
the leading eigenvalue is approximated by $\alpha = \langle k^2\rangle / \langle k\rangle$, where $\langle \cdot \rangle$ represents the average over the $N$ nodes \cite{Chung2003PNAS}. By substituting these relationships in Eq.~\eqref{eq:error-modified-reduction}, we obtain
\begin{equation}
\epsilon_2 \approx \sum_{i=1}^N \left(k_i - \frac{\langle k^2\rangle}{\langle k\rangle}\right)^2 \left( \frac{k_i}{\sum_{\ell=1}^N k_{\ell}} \right)^2 =
\frac{\langle k \rangle^2 \langle k^4 \rangle - 2 \langle k\rangle \langle k^2 \rangle \langle k^3 \rangle + \langle k^2 \rangle^3}
{N \langle k \rangle^4},
\label{eq:epsilon2-config-leading}
\end{equation}
where $\approx$ represents ``approximately equal to''. 
If we instead use a non-leading eigenvector whose associated eigenvalue is of $O(1)$ and assume that $a_i$ and $k_i$ are uncorrelated and that $a_i = O(N^{-1})$, we obtain
\begin{equation}
\epsilon_2 = \sum_{i=1}^N (k_i - \alpha)^2 \cdot O(N^{-2}) = O \left( \langle k^2 \rangle N^{-1}\right).
\label{eq:epsilon2-config-nonleading}
\end{equation}
In degree-heterogeneous networks, the leading term in Eq.~\eqref{eq:epsilon2-config-leading} is $\langle k^4 \rangle \langle k\rangle^{-2} N^{-1}$, which is expected to be much larger than $O(\langle k^2\rangle N^{-1})$ in
Eq.~\eqref{eq:epsilon2-config-nonleading}. Therefore, if $a_i = O(N^{-1})$ for various non-leading eigenvectors, we expect that the spectral method is likely to be optimized by a non-leading eigenvector in degree-heterogeneous networks.

\section{Numerical results}

In this section, we numerically compare the spectral method with the leading eigenvector and that with the error-minimizing eigenvectors.

\subsection{Estimated performance of non-leading eigenvectors for various networks\label{sub:performance-various-networks}}

We first examine the performance of the optimal non-leading eigenvectors in reducing $\epsilon_1$ and $\epsilon_2$ in comparison with the leading eigenvector for several networks.

First, we consider networks with $N$ nodes by the Erd\H{o}s-R\'{e}nyi (ER) random graph in which each node pair is adjacent with probability $\langle k \rangle / (N-1)$; we remind that $\langle k\rangle$ is the mean degree. We generate $200$ networks from the ER random graph 
and use the largest connected component of each network for each pair of $N \in \{ 100, 1000 \}$ and
$\langle k\rangle \in \{4, 10 \}$. For the other network models that we consider in the following text, we also generate $200$ networks and use the largest connected component. The largest connected component of the networks generated by the ER random graph model contains at least 
90\% of the nodes. 
%
%
We then compute the minimizer of $\epsilon_1$ and that of $\epsilon_2$ for each network. 
We find that the leading eigenvector is the minimizer of $\epsilon_1$ and $\epsilon_2$ in all networks. 
Therefore, we should use the spectral method with the leading eigenvector \cite{Laurence2019PhysRevX,Thibeault2020PhysRevResearch} for the ER random graph.

The results are similar for the Watts-Strogatz model of small-world networks \cite{Watts1998Nature}, whereas the leading eigenvector does not minimize $\epsilon_1$ or $\epsilon_2$ for a small fraction of network instances. Specifically, we set the rewiring probability to $0.1$ and generate networks for each pair of $N \in \{ 100, 1000 \}$ and $\langle k\rangle \in \{4, 10 \}$. We have confirmed that the largest connected component of the network always contains all nodes. With $\langle k\rangle = 10$, the leading eigenvector minimizes both $\epsilon_1$ and $\epsilon_2$ for all networks. With $\langle k\rangle = 4$, the minimizer of each type of error (i.e., $\epsilon_1$ or $\epsilon_2$) is not the leading eigenvector only for 1.5\% and 6\% of the networks with $N=100$ and $N=1000$ nodes, respectively. 

\begin{figure}
\includegraphics[width=0.45\textwidth]{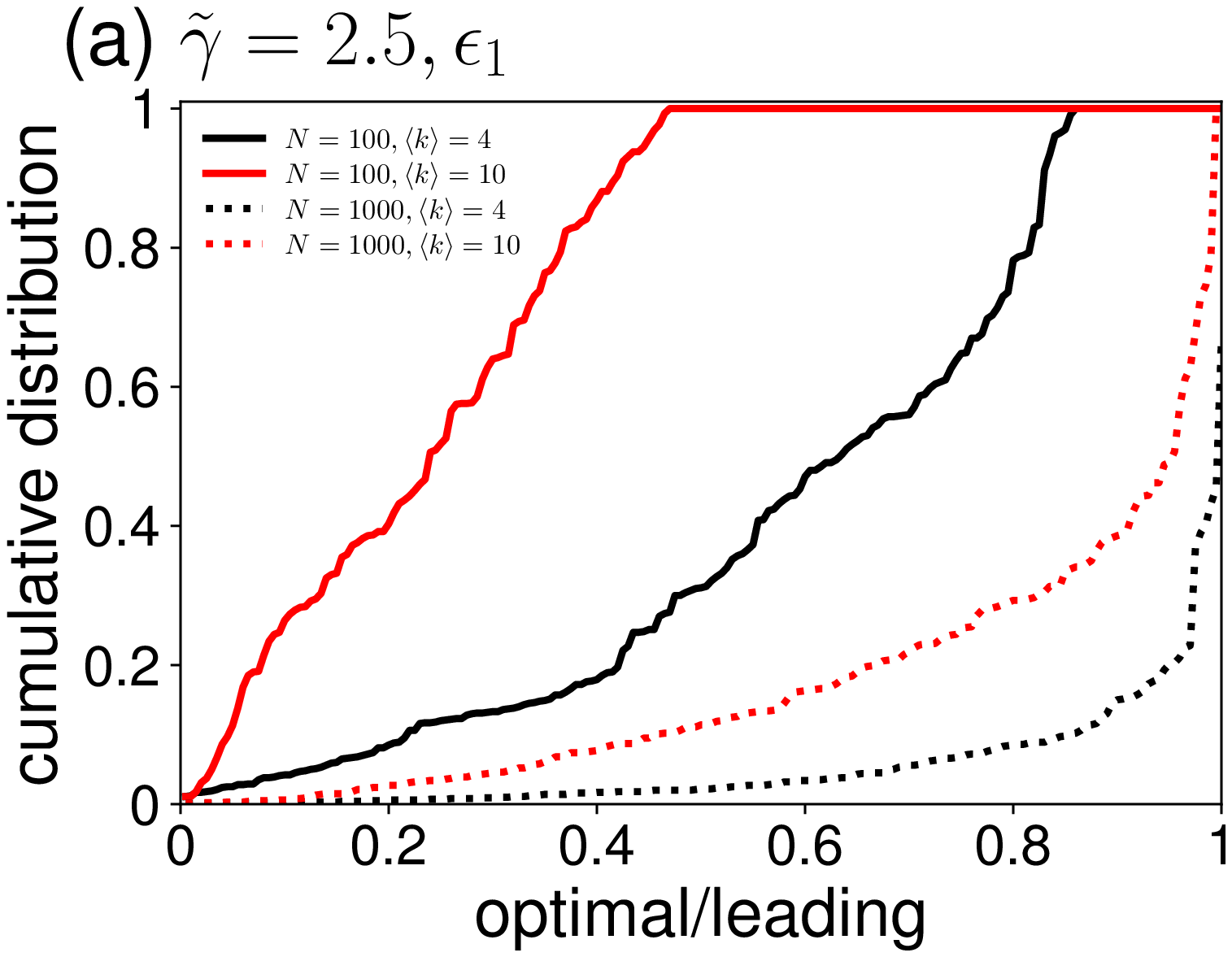}
\includegraphics[width=0.45\textwidth]{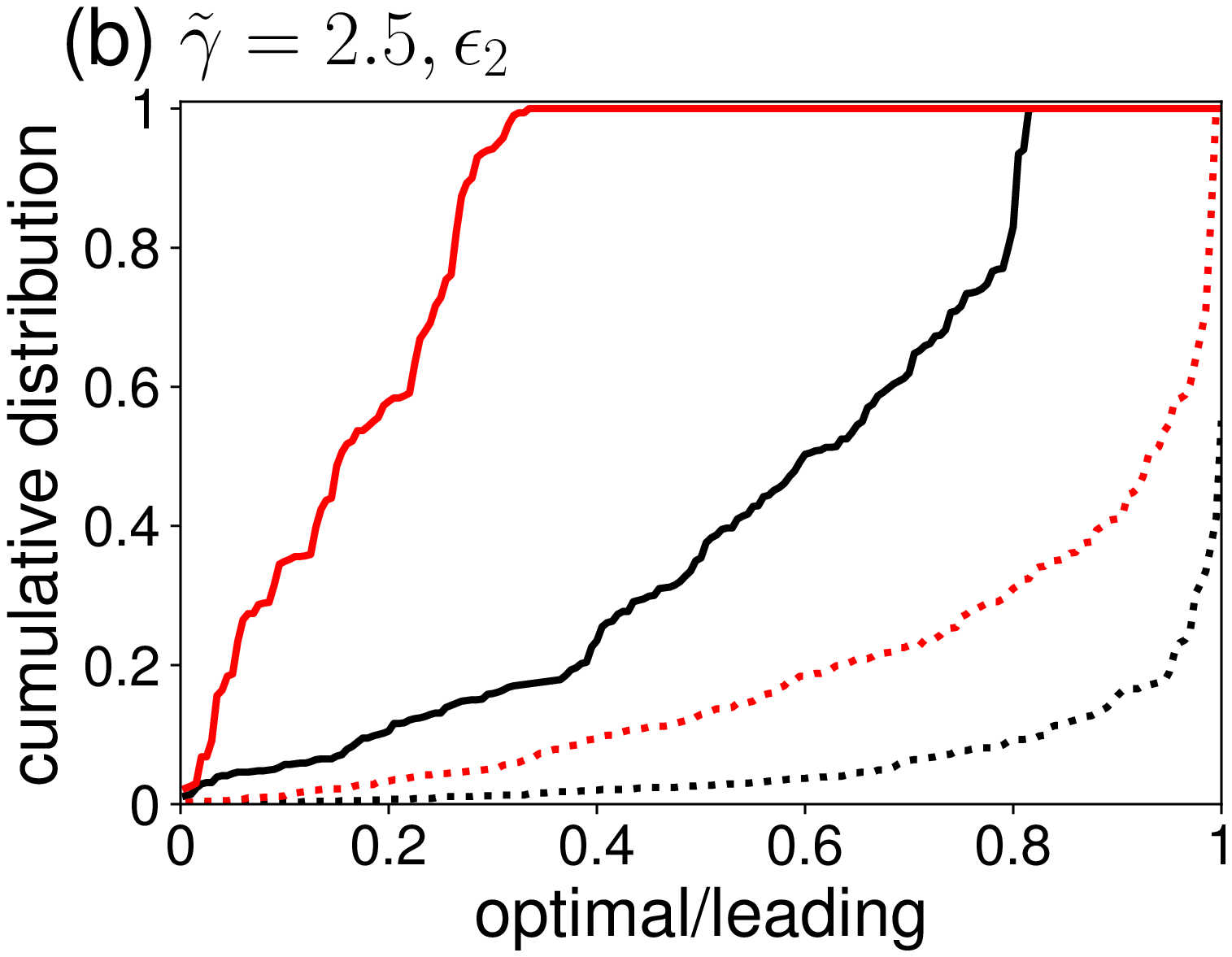}
\includegraphics[width=0.45\textwidth]{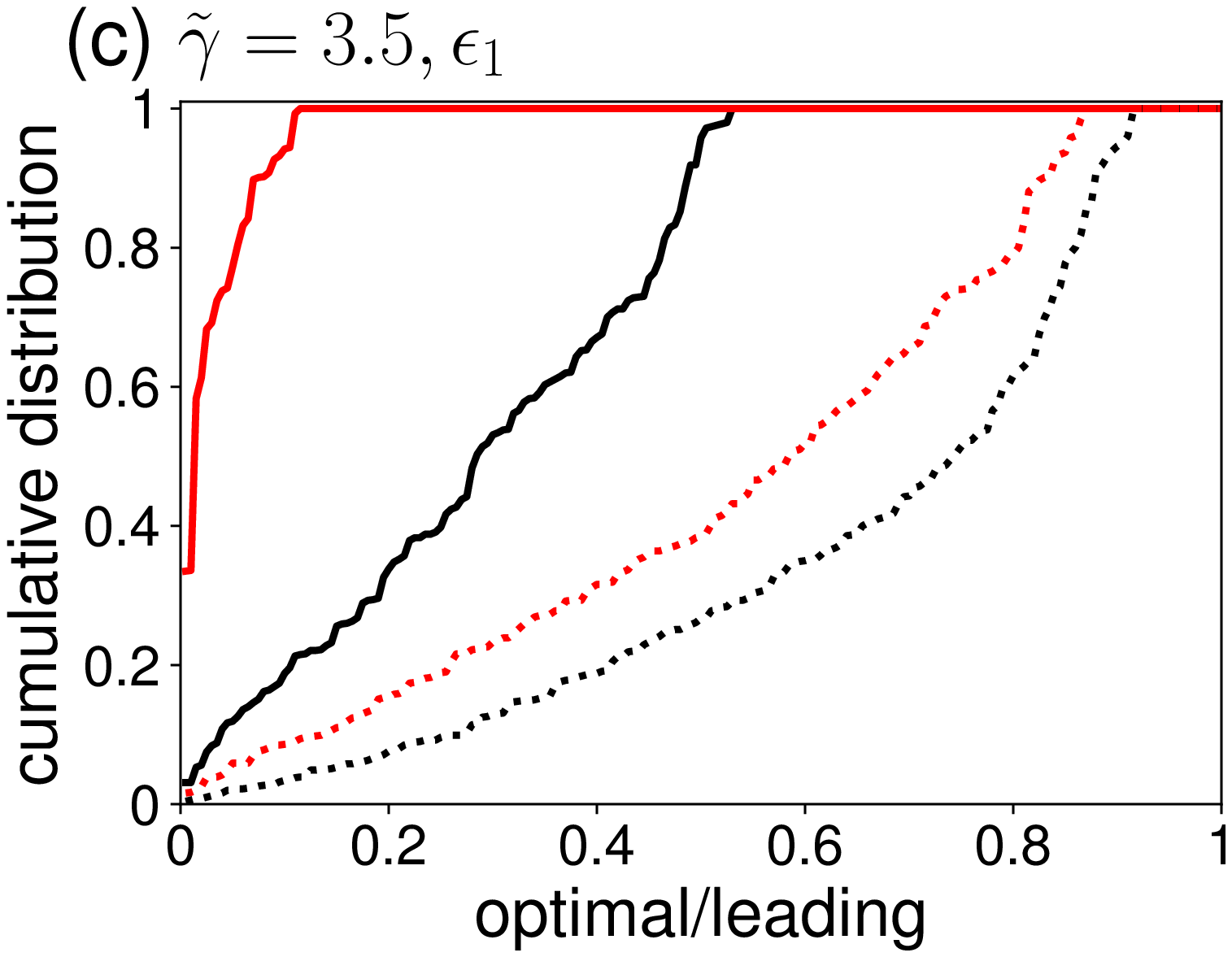}
\includegraphics[width=0.45\textwidth]{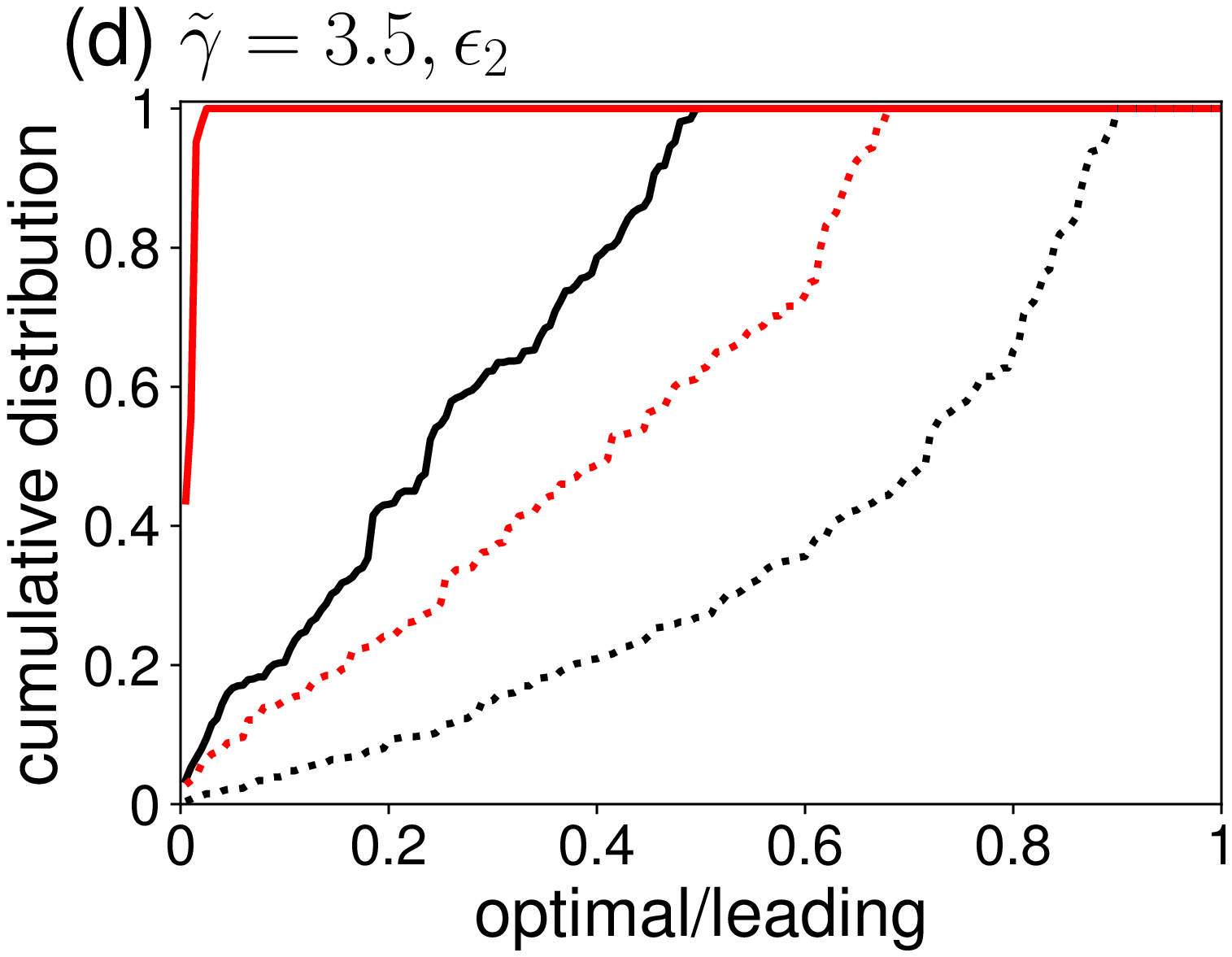}
\includegraphics[width=0.45\textwidth]{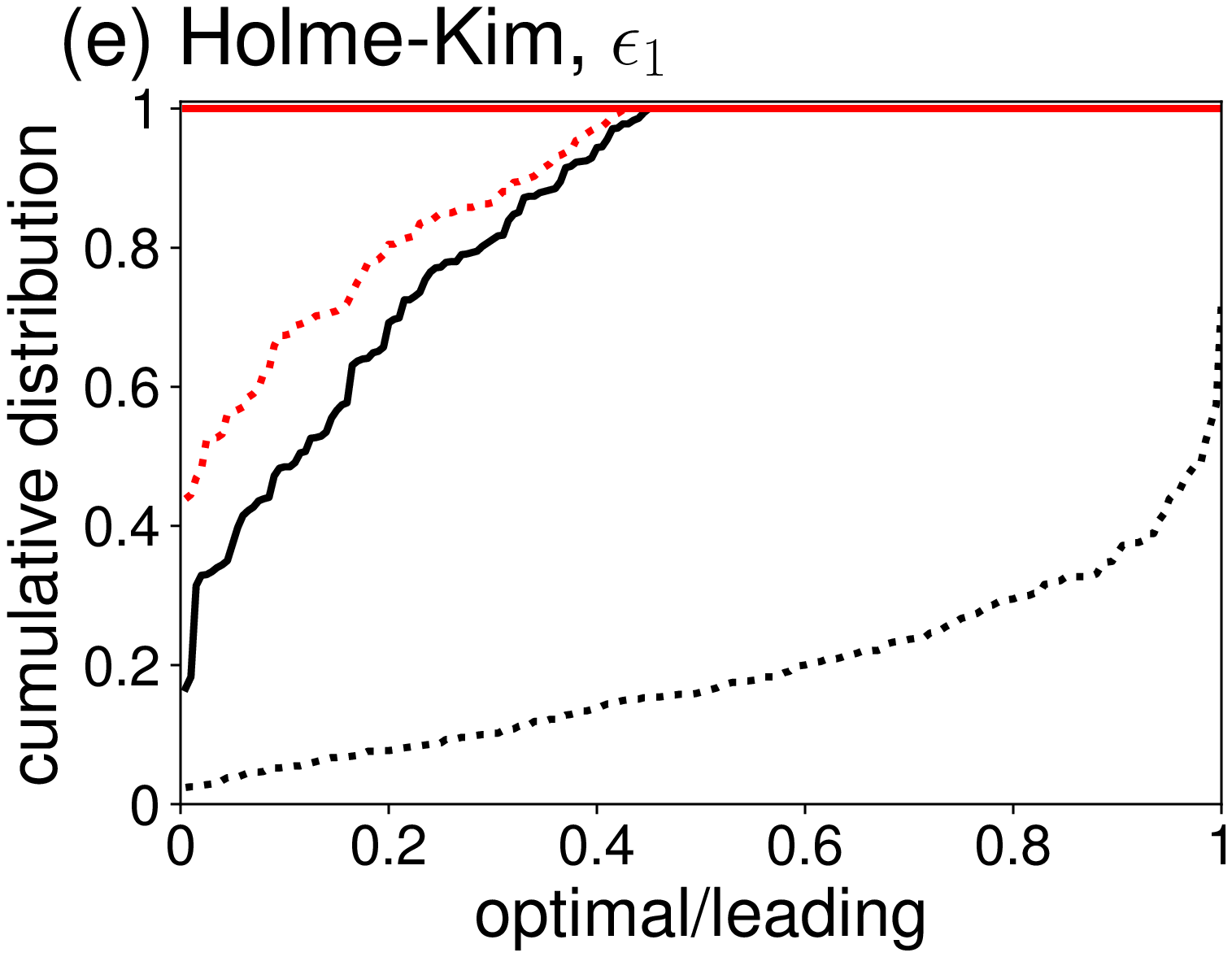}
\includegraphics[width=0.45\textwidth]{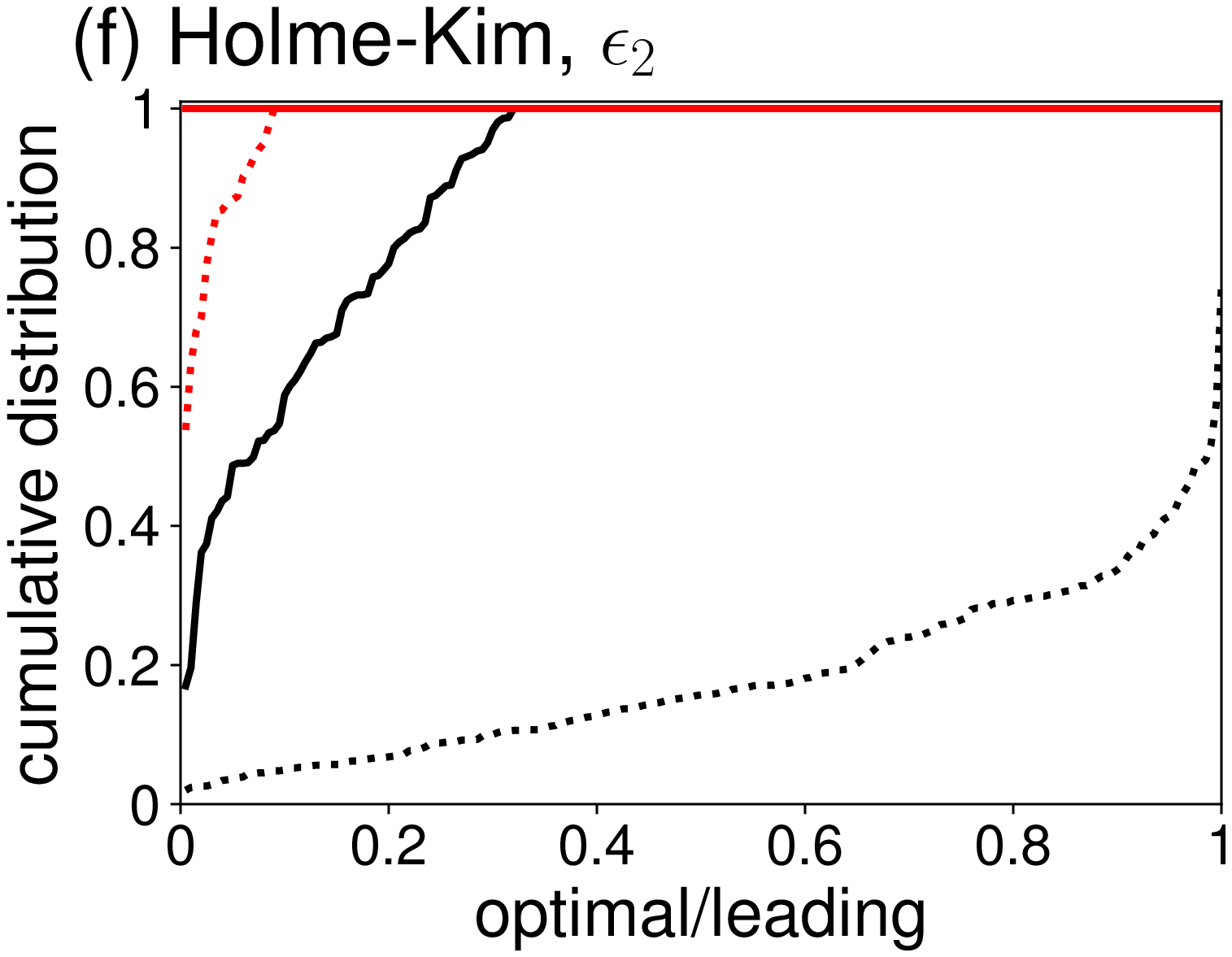}
\caption{Reduction in $\epsilon_1$ and $\epsilon_2$ by their minimizers relative to the case of the leading eigenvector of the adjacency matrix.
We calculate the error (i.e., either $\epsilon_1$ or $\epsilon_2$) for its minimizer divided by the error for the leading eigenvector. The horizontal axis represents this ratio. The vertical axis represents the the cumulative distribution function of this ratio, i.e., the probability that the ratio is larger than the value specified on the horizontal axis. If the cumulative distribution is small across a wide range on the horizontal axis, then the minimizer of $\epsilon_1$ or $\epsilon_2$ tends to be more efficient than the leading eigenvector. (a) Error $\epsilon_1$ for the scale-free networks with $\tilde{\gamma} = 2.5$. (b) Error $\epsilon_2$ for the scale-free networks with $\tilde{\gamma} = 2.5$. (c) Error $\epsilon_1$ for the scale-free networks with $\tilde{\gamma} = 3.5$. (d) Error $\epsilon_2$ for the scale-free networks with $\tilde{\gamma} = 3.5$. (e) Error $\epsilon_1$ for the Holme-Kim model. (f) Error $\epsilon_2$ for the Holme-Kim model.}
\label{fig:ratio}
\end{figure}

Next, we generate networks with power-law degree distributions, which we refer to as scale-free networks, using the configuration model.
To this end, we draw the degree of each node, denoted by $k$, by independently sampling $k$ from a power-law distribution given by $p(k) = \kappa (\tilde{\gamma}-1) / \left[1 + \kappa (k-1) \right]^{\tilde{\gamma}}$, 
%
%
where $k\ge 1$, $\kappa > 0$, and $\tilde{\gamma} > 1$. Because $p(k)$ is a probability density function, we round the sampled $k$ to the nearest integer. The mean of the probability density $p(k)$ is $1 + \left[\kappa(\tilde{\gamma}-2)\right]^{-1}$, we equate it to the mean degree, $\langle k\rangle$.
With the given values of $\langle k \rangle$ and $\tilde{\gamma}$ values, we uniquely determine the $\kappa$ value. We examine combinations of $N \in \{ 100, 1000 \}$, $\langle k\rangle \in \{4, 10\}$, and $\tilde{\gamma} \in \{ 2.5, 3.5 \}$. The largest connected component contains at least 91.5\% of the nodes except when $N=100$, $\langle k \rangle = 4$, and $\tilde{\gamma} \in \{ 2.5, 3.5 \}$, in which case it contains at least 79\% of the nodes (i.e., $79$ nodes).
%
%
%
We calculate the ratio of the $\epsilon_1$ value for its minimizer to 
the $\epsilon_1$ value for the leading eigenvector. By definition, this ratio ranges between $0$ and $1$. If the ratio is small, the minimizer of $\epsilon_1$ is expected to be better at approximating a one-dimensional projection of the original $N$-dimensional dynamics. In contrast, the ratio value equal to $1$ implies that the leading eigenvector minimizes $\epsilon_1$. We similarly measured the ratio of $\epsilon_2$ for its minimizer to $\epsilon_2$ for the leading eigenvector.
For scale-free networks with $\tilde{\gamma} = 2.5$, we show the cumulative distribution of the ratio for $\epsilon_1$ and $\epsilon_2$ in 
in Figs.~\ref{fig:ratio}(a) and~\ref{fig:ratio}(b), respectively. The cumulative distribution shows the fraction of the networks for which the ratio is larger than the value shown on the horizontal axis. Therefore, if the cumulative distribution is small for a range of the ratio value smaller than $1$, then the minimizer of $\epsilon_1$ or $\epsilon_2$ is efficient relative to the leading eigenvector for a large proportion of network instances.
Figures~\ref{fig:ratio}(a) and \ref{fig:ratio}(b) indicate that there are many instances of networks for which the leading eigenvector does not minimize $\epsilon_1$ or $\epsilon_2$ and that the minimization of $\epsilon_1$ or $\epsilon_2$ reduces the error by a large fraction relative to the case of the leading eigenvector in many cases. The results are similar for $\tilde{\gamma} = 3.5$ (see Figs.~\ref{fig:ratio}(c) and \ref{fig:ratio}(d)).
These figures indicate that the minimizer of the error tends to be different from the leading eigenvector and tends to reduce the error by a large amount
when the network is large, sparse, or more heterogeneous in terms of the node's degree (i.e., $\tilde{\gamma} = 2.5$ as opposed to $\tilde{\gamma} = 3.5$). 

Lastly, we investigate the scale-free network model proposed by Holme and Kim, which produces a high clustering coefficient (i.e., many triangles)~\cite{HolmeKim2002PhysRevE-high-clustering}. We set the number of edges that each new node has, denoted by $m$, to $m=2$ and $m=5$ to produce networks whose average degree is approximately equal to $\langle k\rangle = 4$ and $\langle k \rangle = 10$, respectively. We initialize the network by a star graph having $m+1$ nodes.
%
%
%
We set the probability of making a triangle for each added edge to $0.5$. We show the ratio of the minimized error to the error for the leading eigenvector in Figs.~\ref{fig:ratio}(e) and \ref{fig:ratio}(f) for $\epsilon_1$ and $\epsilon_2$, respectively.
The results are qualitatively the same as those for the scale-free networks generated by the configuration model. We also find that
the leading eigenvector tends to minimize the error for more network instances (i.e., the cumulative distribution is equal to 1 for a wider range of the ratio value on the horizontal axis) in the case of the Holme-Kim model than the configuration model, except for $(N, \langle k\rangle) = (1000, 4)$.

\subsection{Networks used in dynamical numerical simulations}

In the following numerical simulations, we focus on $200$ scale-free networks with $N=1000$, $\langle k\rangle \in \{ 4, 10 \}$, and $\tilde{\gamma} = 3.5$.
We also focus on the minimizer of $\epsilon_2$, not that of $\epsilon_1$, because we advocate the minimization of $\epsilon_2$ in the present study (see Section~\ref{sub:our-theory}). To identify the minimizer of $\epsilon_2$ for each network, we discarded the eigenvectors associated with an eigenvalue smaller than $10^{-6}$ including the case of a negative eigenvalue. This exclusion is because, for negative eigenvalues, Eq.~\eqref{eq:modified-reduction-final} behaves qualitatively differently from the case of a positive eigenvalue, and, for positive but tiny eigenvalues, Eq.~\eqref{eq:modified-reduction-final} shows a bifurcation of interest at an extremely large value of the bifurcation parameter. See Section~\ref{sub:tradeoff} for more discussion of this problem. We have found that $162$ out of the $200$ generated networks with $\langle k\rangle = 4$ result in the minimal $e_2$ values that are smaller than $70$\% of the error attained by the leading eigenvector and satisfy the condition that the eigenvalue is at least $10^{-6}$. We have found that $123$ networks meet the same criterion when $\langle k\rangle = 10$. We only use these networks to calculate the statistics in the following analyses because, in those cases, we expect that the spectral method with a non-leading eigenvector may notably be better than that with the leading eigenvector.

In addition to the scale-free networks with $\tilde{\gamma} = 3.5$, we also use a coauthorship network among researchers that published articles on network science by 2006 \cite{Newman2006PhysRevE-collabo}. The original data set contains $1589$ nodes. We only use its largest connected component, which contains $N=379$ nodes and 914 edges. We regard this network as an unweighted network.



\subsection{SIS model\label{sub:SIS}}

First, we consider the deterministic version of the SIS model, which is also called its individual-based approximation \cite{Pastorsatorras2015RevModPhys,KissMillerSimon2017book}, given by 
\begin{eqnarray}
\frac{{\rm d}x_i}{{\rm d}t}= -\mu x_i + \lambda\sum_{j=1}^N w_{ij}(1-x_i)x_j, 
\label{SISeq1}
\end{eqnarray}
where $x_i$ represents the probability that the $i$th node is infectious at time $t$, $\lambda $ is the infection rate, and $\mu$ is the recovery rate. Note that $w_{ij} \in \{ 0, 1 \}$. By definition, each infectious node infects its susceptible neighbor independently at rate $\lambda$. An infectious node independently recovers at rate $\mu$.
Because multiplying a common constant to $\lambda$ and $\mu$ only changes the timescale of the dynamics, we set $\mu=1$ without loss of generality.
For each value of $\lambda$, we run a simulation with the initial condition
$x_i = 0.01, \forall i\in \{1, \ldots, N\}$ until the equilibrium, denoted by $\bm x^* = (x_1^*, \ldots, x_N^*)$, is sufficiently closely reached.

\begin{figure}
\includegraphics[height=!,width=0.95\textwidth]{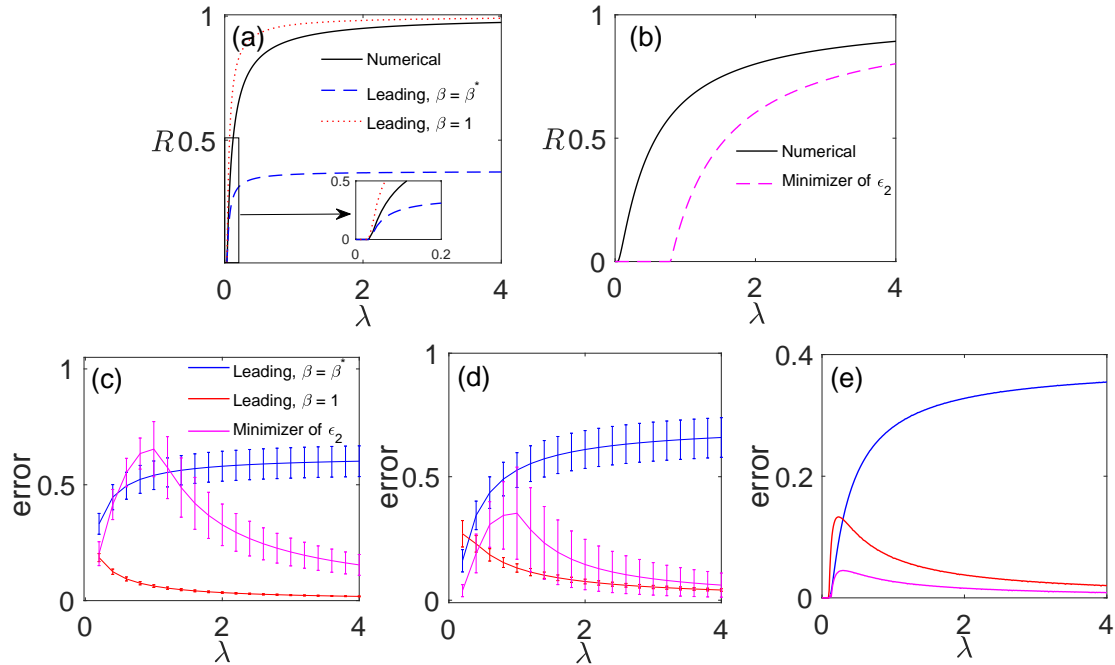}
\caption{Results for the SIS model. (a) Bifurcation diagram in terms of the observable, $R$, using the spectral method with the leading eigenvector as $\bm a$ for a scale-free network with $N=1000$, $\langle k \rangle=10$, and $\tilde{\gamma} =3.5$. The inset is a magnification near $\lambda=0$ to show the behavior near the epidemic threshold. (b) Bifurcation diagram for the same network when $\bm a$ is the minimizer of $\epsilon_2$. (c) Error for the spectral method with the different $\bm a$ and $\beta$ values for scale-free networks with $N=1000$, $\langle k \rangle=10$, and $\tilde{\gamma} = 3.5$. We have calculated the error bars, which represent the standard deviations, on the basis of the 123 networks of which the minimized error $\epsilon_2$ is less than 70\% of the error obtained with the leading eigenvector. (d) Error for scale-free networks
with $N=1000$, $\langle k \rangle = 4$, and $\tilde{\gamma} = 3.5$. We have calculated the error bars on the basis of the 162 networks of which the minimized error $\epsilon_2$ is less than 70\% of the error obtained with the leading eigenvector. (e) Error for the coauthorship network.}
\label{fig:SIS}
\end{figure}

We show the bifurcation diagram for observable $R = \bm a^{\top} \bm x^*$, where $\bm a$ is the leading eigenvector of the adjacency matrix, as a function of the infection rate, $\lambda$, for a scale-free network with $N=1000$ nodes, $\langle k \rangle = 10$, and $\tilde{\gamma} = 3.5$ in Fig.~\ref{fig:SIS}(a).
Note that $R$ is a weighted fraction of infectious nodes in the equilibrium.
The solid line represents the results obtained from direct numerical simulations of the model.
The dashed line represents the bifurcation diagram for the original spectral method, i.e., with the leading eigenvector as the weight vector $\bm a$ and $\beta = \beta^*$ (see Eq.~\eqref{eq:Laurence-final}). We observe that the spectral method is accurate at locating the bifurcation point.
However, the spectral method considerably underestimates $R$ in the endemic phase (i.e., for $\lambda$ values larger than the epidemic threshold).
The dotted line in Fig.~\ref{fig:SIS}(a) shows the approximation of the same observable, $R$, by the spectral method in which we force $\beta=1$, corresponding to Eq.~\eqref{eq:modified-reduction-final}, and continue to use the leading eigenvector as $\bm a$. The use of $\beta=1$ is discussed in Ref.~\cite{Laurence2019PhysRevX}. We find that this one-dimensional reduction also locates the bifurcation point of the original dynamical system accurately and that it is better at approximating the $R$ value in the endemic phase than with $\beta = \beta^*$. We compare the numerical results and the spectral method in which $\bm a$ is the minimizer of $\epsilon_2$ in Fig.~\ref{fig:SIS}(b). Note that the observable $R$ is now different from that used in Fig.~\ref{fig:SIS}(a) because we have changed $\bm a$.
Figure~\ref{fig:SIS}(b) indicates that the spectral method with the minimizer of $\epsilon_2$ is worse than that with the leading eigenvector at accurately locating the bifurcation point. The minimizer of $\epsilon_2$ is better at approximating $R$ than the leading eigenvector combined with $\beta = \beta^*$ but worse than the leading eigenvector combined with $\beta = 1$. The reason why the spectral method with the combination of the leading eigenvector as $\bm a$ and $\beta=1$ does not minimize $\epsilon_2$ but works better than the minimizer of $\epsilon_2$ is unclear. It may be because higher-order terms in terms of $\Delta x_i$ in our theory are nonnegligible or because the range of $D$ values we have explored is not sufficiently far from the epidemic threshold estimated by the spectral method using the minimizer of $\epsilon_2$.

We compare the error among the three methods in Fig.~\ref{fig:SIS}(c). At each $\lambda$ value, we defined the error as the absolute value of the difference between the $R$ obtained by the direct numerical simulation and that obtained from the one-dimensional reduction, which we divided by the $R$ value at the largest $\lambda$ value, i.e., $\lambda = 4$. We normalized the error in this manner because the true value of the observable $R$ depends on $\bm a$. We use the absolute error instead of the relative error because $R$ is close to or equal to 0 when $\lambda$ is small. The error bars represent the mean and standard deviation. The figure confirms that the approximation error is the smallest in the case of the spectral method with the combination of the leading eigenvector and $\beta=1$, the second smallest in the case of the minimizer of $\epsilon_2$, and the largest in the case of the combination of the leading eigenvector and $\beta = \beta^*$, when $\lambda$ is sufficiently larger than the epidemic threshold for the one-dimensional reduction with the minimizer of $\epsilon_2$ (i.e., $\lambda \approx 0.80$). The results are qualitatively the same for scale-free networks with $\langle k\rangle = 4$, as we show in Fig.~\ref{fig:SIS}(d).
We show the approximation error for the coauthorship network in Fig.~\ref{fig:SIS}(e). For this network, the minimizer of $\epsilon_2$ realizes a smaller error than the leading eigenvector combined with $\beta = \beta^*$ or $\beta = 1$ for all values of $\lambda$. It should also be noted that the spectral method with the minimizer of $\epsilon_2$ accurately locates the epidemic threshold for the coauthorship network.

\subsection{Double-well system}

In this section, we consider the coupled double-well system given by 
\begin{equation}
\frac{{\rm d}x_i}{{\rm d}t}=-\left(x_i-r_1\right)\left(x_i-r_2\right)\left(x_i-r_3\right) +D\sum_{j=1}^N w_{ij} x_j, 
\label{double_well_system1}
\end{equation} 
where $x_i$ is the state of the node $i$; $D$ is the coupling strength, which we assume to be common for all edges; $r_1$, $r_2$, and $r_3$ are constants satisfying $r_1 < r_2 < r_3$~\cite{Wunderling_Chaos2020, Brummitt_JRSI2015, Kronke_PRE2020, Wunderling_NJP2020, Klose_RSOS2020, Wunderling_ESD2021}. We set $r_1 = 1$, $r_2 =2 $, and $r_3 =5 $. When the coupling is absent, the dynamics is bistable, with $x_i = r_1$ and $x_i = r_3$ being stable equilibria and $x_i = r_2$ being the unique unstable equilibrium.
We use the initial condition $x_i = 0.01, \forall i\in \{1, \ldots, N\}$ for each value of $D$. With this initial condition, $x_i$ will converge to the lower equilibrium, i.e., $r_1$, if there is no coupling.

\begin{figure}
\includegraphics[height=!,width=0.95\textwidth]{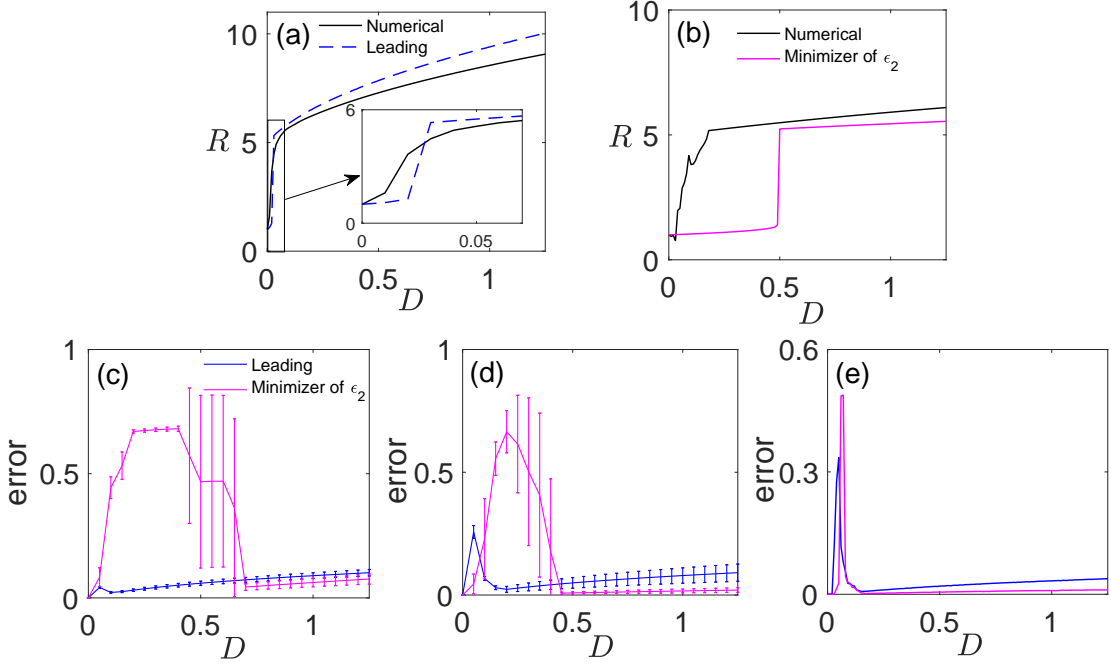}
\caption{Results for the double-well system. (a) Bifurcation diagram in terms of $R=\bm a^{\top} \bm x^*$, where $\bm a$ is the leading eigenvector, for the same scale-free network with $N=1000$, $\langle k \rangle=10$, and $\tilde{\gamma} =3.5$ as the one used in Fig.~\ref{fig:SIS}. (b) Bifurcation diagram for the same network when $\bm a$ is the minimizer of $\epsilon_2$. (c) Error compared between the leading eigenvector and the minimizer of $\epsilon_2$ as $\bm a$ for scale-free networks with $N=1000$, $\langle k \rangle=10$, and $\tilde{\gamma} = 3.5$. (d) Error for scale-free networks with $N=1000$, $\langle k \rangle = 4$, and $\tilde{\gamma} = 3.5$. (e) Error for the coauthorship network.}
\label{fig:double-well}
\end{figure}

We plot in Fig.~\ref{fig:double-well}(a) $R = \bm a^{\top} \bm x^*$, where $\bm a$ is the leading eigenvector, against $D$
for the coupled double-well system on the same scale-free network with $N=1000$, $\langle k\rangle = 10$, and $\tilde{\gamma} = 3.5$ as the one used in Figs.~\ref{fig:SIS}(a) and \ref{fig:SIS}(b).
The theoretical estimate of $R$ shown in Fig.~\ref{fig:double-well}(a) does not depend on the $\beta$ value because function $G(x_i, x_j)$ does not depend on $x_i$ for the double-well system such that Eq.~\eqref{eq:Laurence-final} does not depend on $\beta$.
We show in Fig.~\ref{fig:double-well}(b) the same relationships when $\bm a$ is the minimizer of $\epsilon_2$.
Similar to the case of the SIS model, the spectral method with the leading eigenvector is substantially better at locating the bifurcation point than the minimizer of $\epsilon_2$ is. The accuracy at approximating $R$ is slightly better for the minimizer of $\epsilon_2$ than the leading eigenvector except near the bifurcation point, which we confirm by the statistical analysis of the absolute error shown in Fig.~\ref{fig:double-well}(c). The results are similar for $\langle k\rangle = 4$ except that the minimizer of $\epsilon_2$ yields significantly smaller errors than the leading eigenvector when $D$ is large (see Fig.~\ref{fig:double-well}(d)). The results are similar for the coauthorship network (see Fig.~\ref{fig:double-well}(e)). We will discuss the relatively low accuracy near the bifurcation point in Section~\ref{sub:tradeoff}.

\subsection{Generalized Lotka-Volterra dynamics}

Third, we consider the generalized Lotka-Volterra (GLV) dynamics is given by
\begin{eqnarray}
\frac{{\rm d}x_i}{{\rm d}t} = \overline{\lambda} x_i +D\sum_{j=1}^N w_{ij} x_i x_j,
\label{GLVeq1}
\end{eqnarray}
where $x_i$ represents the abundance of the $i$th species, $\overline{\lambda}$ is the intrinsic growth rate of the species, and $D$ is the coupling strength as in the case of the double-well system \cite{Tu2017PhysRevE, Tu_IScience2021}. The nontrivial equilibrium of this dynamical system is given by $\bm x^*= - \overline{\lambda} W^{-1} \bm 1$, where we remind that $W=(w_{ij})$ is the adjacency matrix. This equilibrium is globally asymptotically stable if and only if $W$ is negative definite \cite{Grilli_NatComm2017}. Therefore, only for the GLV dynamics, we change the definition of the adjacency matrix to set $w_{ii}=-(\alpha_{\max}+1)/D$, $\forall i\in \{1, \ldots, N \}$, where $\alpha_{\max}$ is the largest eigenvalue of $W$, which makes $W$ negative definite \cite{KunduKoriMasuda2022PhysRevE}. With this $w_{ii}$, we rewrite Eq.~\eqref{GLVeq1} as 
\begin{equation}
\frac{{\rm d}x_i}{{\rm d}t} = \overline{\lambda} x_i -cx_i^2+D\sum_{j=1; j\neq i}^N w_{ij}x_ix_j,
\label{GLVeq11} 
\end{equation}
where $c = \alpha_{\max} + 1$. Therefore, we set $F(x_i) = \overline{\lambda} x_i - cx_i^2$ and $G(x_i, x_j) = D x_i x_j$ \cite{KunduKoriMasuda2022PhysRevE}.

The equilibrium of Eq.~\eqref{GLVeq11} is given by  $\bm x^* = - \overline{\lambda}(DW - cI)^{-1} \bm 1$, where $W$ is the weighted adjacency matrix of the original network with the diagonal entries being equal to 0, and $I$ is the $N\times N$ identity matrix. Therefore, we obtain
\begin{align}
R =& \bm a^{\top} \bm x^*\notag\\
=& - \overline{\lambda} \bm a^{\top} (DW - cI)^{-1} \bm 1\notag\\
=& - \overline{\lambda} \bm a^{\top} (D\alpha-c)^{-1} \bm 1\notag\\
=& \frac{\overline{\lambda}}{\alpha_{\max} + 1 - D \alpha}.
\label{eq:R-GLV-full-analytical}
\end{align}
On the other hand, the $R$ value in the equilibrium for the one-dimensional reduction, Eq.~\eqref{eq:Laurence-final}, including the case of our modified spectral method (i.e., 
Eq.~\eqref{eq:modified-reduction-final}) with the replacement of $\beta^*$ by $1$, is given by
\begin{equation}
R = \frac{\overline{\lambda}}
{\alpha_{\max}+1 - \beta^* D \alpha}.
\end{equation}
Therefore, the spectral method is exact if and only if $\beta^* = 1$ and regardless of which eigenvector of $W^{\top}$ we use as $\bm a$.

\begin{figure}
\includegraphics[height=!,width=0.95\textwidth]{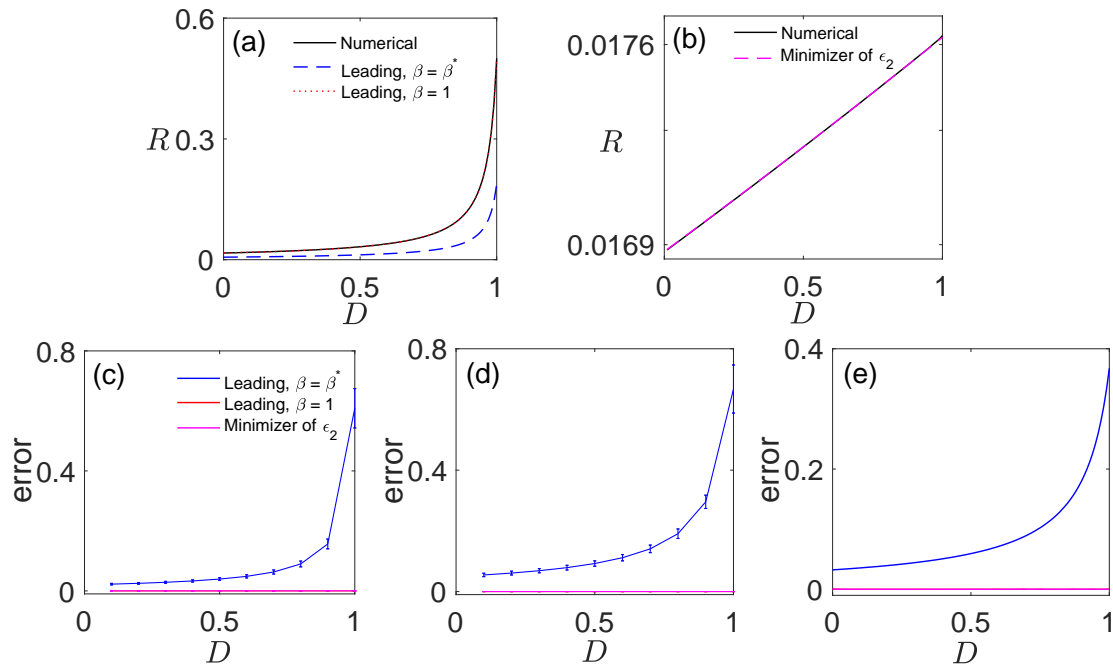}
\caption{Results for the GLV model. (a) Relationship between $R$ and $D$ for the spectral method with the leading eigenvector as $\bm a$ for the same scale-free network with $N=1000$, $\langle k \rangle=10$, and $\tilde{\gamma} =3.5$ as the one used in Fig.~\ref{fig:SIS}. The theoretical estimate for $\beta=1$ shown by the dotted line completely overlaps the numerical results shown by the solid line. (b) Same when we use the minimizer of $\epsilon_2$ as $\bm a$. The theoretical estimate by the dashed line completely overlaps the numerical results shown by the solid line. (c) Error compared among the three cases for scale-free networks with $N=1000$, $\langle k \rangle=10$, and $\tilde{\gamma} = 3.5$. (d) Error for scale-free networks with $N=1000$, $\langle k \rangle = 4$, and $\tilde{\gamma} = 3.5$. (e) Error for the coauthorship network. In (c), (d), and (e), the error in the case of the leading eigenvector combined with $\beta=1$ and the minimizer of $\epsilon_2$ is equal to $0$ regardless of the $D$ value.} 
\label{fig:GLV}
\end{figure}

With $\overline{\lambda} = 0.5$, we show in Fig.~\ref{fig:GLV} the $R$ values and the error as a function of $D$ for the GLV model on the different networks. For Figs.~\ref{fig:GLV}(a) and \ref{fig:GLV}(b), we use the same scale-free network with $N=1000$, $\langle k\rangle = 10$, and $\tilde{\gamma} = 3.5$ as the one used in Figs.~\ref{fig:SIS}(a) and \ref{fig:SIS}(b). With the leading eigenvector as $\bm a$ and $\beta = \beta^*$, the spectral method is not accurate at approximating $R$ (see the dashed line in Fig.~\ref{fig:GLV}(a)). In contrast, as the theory predicts, the theoretically obtained $R$ perfectly matches the numerically obtained $R$ when one combines the leading eigenvector and $\beta = 1$ (see the dotted line in Fig.~\ref{fig:GLV}(a), which completely overlaps the solid line) or uses the minimizer of $\epsilon_2$ (see Fig.~\ref{fig:GLV}(b), in which the dashed line completely overlaps the solid line). Note that $R$ as a function of $D$ looks qualitatively different between Figs.~\ref{fig:GLV}(a) and \ref{fig:GLV}(b).
This is because Eq.~\eqref{eq:R-GLV-full-analytical} diverges to infinity at $D = D_{\rm c} \equiv (\alpha_{\max}+1)/\alpha$ as one increases $D$ from a small value, and
$D\approx 1$, which is the largest value of $D$ in these figures, is much closer to $D_{\rm c}$ when $\alpha = \alpha_{\max}$ than when $\alpha$ is the eigenvalue associated with the minimizer of $\epsilon_2$. We statistically confirm the results shown in Figs.~\ref{fig:GLV}(a) and \ref{fig:GLV}(b) for various scale-free networks in Figs.~\ref{fig:GLV}(c) and \ref{fig:GLV}(d) and the coauthorship network in Fig.~\ref{fig:GLV}(e).

\section{Advantages of the leading eigenvector to the minimizer of $\epsilon_2$}

The theoretical and numerical results shown in the previous sections suggest that the spectral method using the minimizer of $\epsilon_2$ as $\bm a$ outperforms that using the leading eigenvector as $\bm a$ in many cases. However, in this section, we discuss two reasons why we still prefer the leading eigenvector as $\bm a$, but combined with $\beta = 1$, in the spectral method.

\subsection{The spectral method with the leading eigenvector better predicts the bifurcation point \label{sub:tradeoff}}

For the SIS model and double-well system, we have found that the spectral method using the minimizer of $\epsilon_2$ as $\bm a$ is much less accurate at locating the bifurcation point than the spectral method using the leading eigenvector as $\bm a$, as shown in Figs.~\ref{fig:SIS}(a), \ref{fig:SIS}(b), \ref{fig:double-well}(a), and \ref{fig:double-well}(b). The reason for this phenomenon is as follows.

The modified spectral method (i.e., Eq.~\eqref{eq:modified-reduction-final}) as well as the original spectral method (i.e., Eq.~\eqref{eq:Laurence-final}) with $\beta^*$ being replaced by $1$ (see Figs.~\ref{fig:SIS}, \ref{fig:double-well}, and \ref{fig:GLV}) uses the eigenvalue of the adjacency matrix, $\alpha$, as the effective coupling strength.
We can interpret the earlier one-dimensional reduction developed by Gao \text{et al.} \cite{GaoBarzelBarabasi2016Nature} as a mean-field theory to approximate the leading eigenvalue $\alpha$ by a function of the nodes' degrees. 
Suppose that $G(x_i, x_j) = D \tilde{G}(x_i, x_j)$, where $D$ is the coupling strength, which we regard as the bifurcation parameter. Note that, in the SIS model, the infection rate, $\lambda$, plays the role of $D$.
Assume that the one-dimensional dynamical system given by Eq.~\eqref{eq:modified-reduction-final} and $G(x_i, x_j) = D \tilde{G}(x_i, x_j)$ with $\alpha$ being the leading eigenvalue (i.e., $\alpha_{\max}$) undergoes a bifurcation at $D = D_{\rm c, \max}$. We denote by $D_{\rm c, org}$ the value of $D$ at which the bifurcation occurs in the original $N$-dimensional dynamical system. Empirically, the spectral method using the leading eigenvector as $\bm a$
is not necessarily accurate at anticipating the bifurcation point, and one tends to obtain
\begin{equation}
\Delta D_{{\rm c}, \max} \equiv D_{{\rm c}, \max} - D_{\rm c, org} > 0
\label{eq:Delta-Dc}
\end{equation}
such that the spectral method tends to overestimate the bifurcation point for some models of population dynamics \cite{KunduKoriMasuda2022PhysRevE}. 

If we use a non-leading eigenvector as $\bm a$, we only replace $\alpha$ in Eq.~\eqref{eq:modified-reduction-final} by the eigenvalue associated with $\bm a$. Therefore, the bifurcation in the one-dimensional reduced dynamical system occurs at $D = D_{\rm c, NL}$, where
\begin{equation}
\alpha_{\rm NL} D_{\rm c, NL} = \alpha_{\max} D_{{\rm c}, \max},
\label{eq:alpha-Dc}
\end{equation}
and $\alpha_{\rm NL}$ is the non-leading eigenvalue associated with the eigenvector $\bm a$ under consideration.
Using Eqs.~\eqref{eq:Delta-Dc} and \eqref{eq:alpha-Dc}, we evaluate the error in locating the bifurcation point by the spectral method with a non-leading eigenvector as follows:
\begin{equation}
\Delta D_{\rm c, NL} \equiv D_{\rm c, NL} - D_{\rm c, org} = \frac{\alpha_{\max}}{\alpha_{\rm NL}} D_{{\rm c}, \max} - D_{\rm c, org}.
\label{eq:farther-bifurcation-point}
\end{equation}
We find that $\Delta D_{\rm c, NL} > \Delta D_{{\rm c}, \max} > 0$ and that the difference between $\Delta D_{\rm c, NL}$ and $\Delta D_{{\rm c}, \max}$ is large if 
the eigenvalue ratio, $\alpha_{\max} / \alpha_{\rm NL}$, is large. Therefore, if the minimizer of $\epsilon_2$ is associated with a small eigenvalue of the adjacency matrix, the bifurcation point for the one-dimensional dynamical system, $D = D_{\rm c, NL}$, tends to be much larger than $D_{\rm c, org}$. Then, for $D \in (D_{\rm c, org}, D_{\rm c, NL})$, the one-dimensional reduction is qualitatively wrong at describing the original $N$-dimensional dynamical system such that our method is not expected to be accurate at approximating $R$ for the original dynamical system.
This is why we excluded in the beginning of Section~\ref{sub:SIS} the eigenvectors $\bm a$ that are
associated with positive eigenvalues with tiny magnitudes and negative eigenvalues from the candidates of the minimizer of the error.

\begin{figure}
\includegraphics[width=0.7\textwidth]{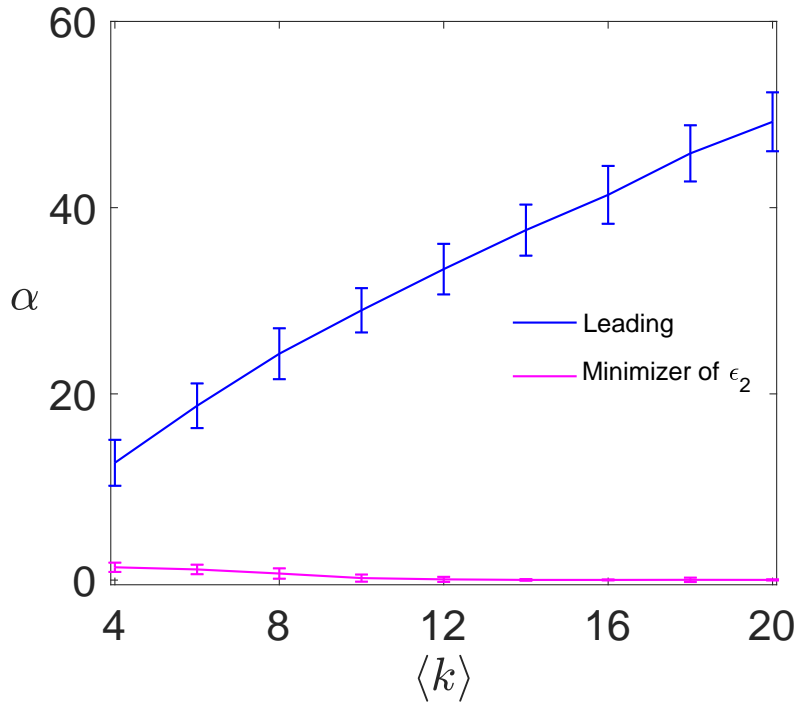}
\caption{Eigenvalues of the adjacency matrix of scale-free networks with $N=1000$ nodes and $\tilde{\gamma} = 3.5$.
We show the leading eigenvalue and the eigenvalue associated with the minimizer of $\epsilon_2$. For each $\langle k\rangle$ value, the error bar represents the average and standard deviation calculated on the basis of the subset of $200$ networks for which the eigenvalue associated with the eigenvector minimizing $\epsilon_2$ is at least $10^{-6}$.} 
\label{fig:degree_eval}
\end{figure}

To further demonstrate the relevance of this reasoning, we show in Fig.~\ref{fig:degree_eval} the largest eigenvalue and the eigenvalue whose associated eigenvector minimizes $\epsilon_2$. We use scale-free networks with $N = 1000$ and $\tilde{\gamma} = 3.5$, and vary $\langle k\rangle$. 
As we have done in the numerical simulations whose results are shown in Figs.~\ref{fig:SIS}, \ref{fig:double-well}, and \ref{fig:GLV}, we exclude the eigenvalues smaller than $10^{-6}$. Figure~\ref{fig:degree_eval} indicates that the leading eigenvalue increases as $\langle k\rangle$ increases, 
as theory predicts \cite{Chung2003PNAS}, and that the eigenvalue associated with the minimizer of $\epsilon_2$ decreases as $\langle k\rangle$ increases. Therefore, Eq.~\eqref{eq:farther-bifurcation-point} implies that $D_{\rm c, NL}$ tends to be much larger than $D_{\rm c, org}$ as $\langle k\rangle$ increases.

\subsection{The spectral method with the leading eigenvector is more robust against noise \label{sub:noise}}

We expect that the spectral method using the leading eigenvector as $\bm a$ is more robust against noise than that using the minimizer of $\epsilon_2$ for the following reason.
Consider dynamical noise added to our dynamical system on networks, i.e., Eq.~\eqref{eq:system}. Assume that each $x_i$ fluctuates around the equilibrium in the case without noise, $x_i^*$, with mean $0$ and standard deviation $\sigma_i$. For simplicity, we also assume that $x_i$ and $x_j$, where $i\neq j$, are uncorrelated in the equilibrium; note that this assumption is just for the sake of discussion here and does not hold in general due to the interaction between different nodes via edges \cite{Risken1989book,ChenOdea2019SciRep}.
Then, the expectation of $R$ in the equilibrium is given by $R = \sum_{i=1}^N a_i x_i^*$. 
The standard deviation of $R$, denoted by $\sigma_R$, is given by $\sigma_R = \sqrt{\sum_{i=1}^N \left| a_i \right| \sigma_i^2}$. 
We remind that $\bm a$ is normalized such that $\sum_{i=1}^N a_i = 1$.
We combine the following two observations to argue that $R$ using the leading eigenvector as $\bm a$ is more robust against noise than $R$ using the minimizer of $\epsilon_2$. First, the expectation of $R$ depends on $\bm a$ unless $x_1 = \cdots = x_N$.
In our numerical results shown in Figs.~\ref{fig:SIS}(a), \ref{fig:SIS}(b), \ref{fig:double-well}(a), \ref{fig:double-well}(b), \ref{fig:GLV}(a), and \ref{fig:GLV}(b), 
$R$ tends to be smaller when $\bm a$ is the minimizer of $\epsilon_2$ than when $\bm a$ is the leading eigenvector.
Second, and more importantly, $\sigma_R$ would be larger with the minimizer of $\epsilon_2$ than with the leading eigenvector.
When $\bm a$ is the leading eigenvector, the Perron-Frobenius theorem guarantees that $a_i > 0$, $\forall i\in \{1, \ldots, N\}$. In contrast, when $\bm a$ is a nonleading eigenvector, including the case of the minimizer of $\epsilon_2$, the orthogonality of the eigenvectors associated with the different eigenvalues implies that some of $a_i$ are positive and others are negative. Then, because $\sum_{i=1}^N a_i = 1$, the $\left| a_i \right|$ value tends to be larger for a nonleading eigenvector than the leading eigenvector. For example, suppose that $N=3$, that the normalized leading eigenvector is $\bm a = (1/2, 1/4, 1/4)^{\top}$, and a normalized nonleading eigenvector is $\bm a = (-1, 1, 1)^{\top}$. Then, if $\sigma_1 = \sigma_2 = \sigma_3 = \sigma$, one obtains $\sigma_R = \sqrt{\sum_{i=1}^N \left| a_i \right|} \sigma = \sigma$ for the leading eigenvector and $\sigma_R =  \sqrt{3}\sigma$ for the nonleading eigenvector. Therefore, the signal-to-noise ratio for $R$ in the case of the leading eigenvector, quantified by the expectation divided by the standard deviation of $R$, should be smaller than the same ratio in the case of a non-leading eigenvector.

To examine the relevance of this argument, we run numerical simulations of the three dynamical systems with noise. We add an independent white noise with the intensity $\sqrt{\mathcal{D}}$  to the right-hand side of each of the $N$ differential equations constituting the dynamical system on the network. 
In other words, we add a value sampled from the Gaussian distribution with mean $0$ and standard deviation $\sqrt{\mathcal{D}{\rm d}t}$ to the right-hand side of the differential equation in each integration time step of size ${\rm d}t$.
We set $\sqrt{\mathcal{D}} = 0.2$ for the SIS and GLV models and $\sqrt{\mathcal{D}} = 10$ for the double-well system.
For the SIS and GLV models, once $x_i$ becomes negative due to the added noise, it tends to diverge to $-\infty$. To prevent this phenomenon, once any $x_i$ becomes negative in any simulation time step, we reset $x_i$ to $10^{-6}$. 

\begin{figure}
\includegraphics[width=\textwidth]{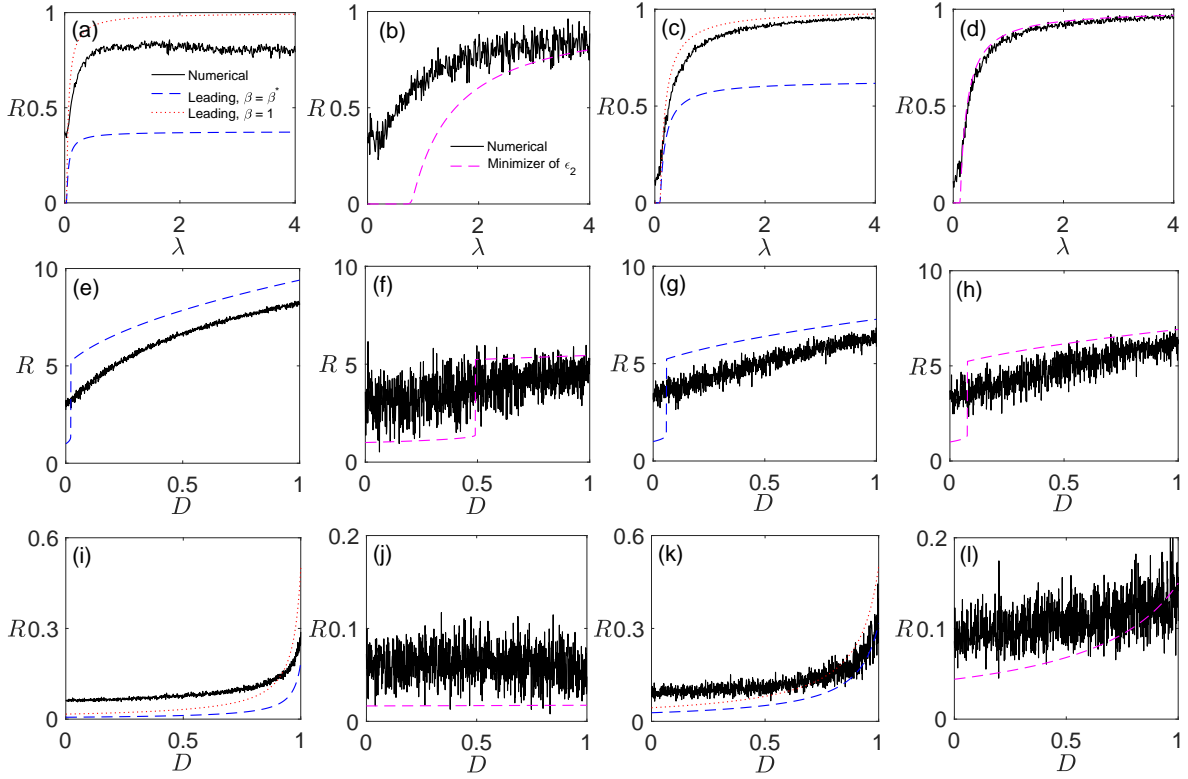}
\caption{$R$ in the presence of noise. (a) SIS model on the scale-free network; $\bm a$ is the leading eigenvector. (b) SIS model on the scale-free network; $\bm a$ is the minimizer of $\epsilon_2$. (c) SIS model on the coauthorship network; $\bm a$ is the leading eigenvector. (d) SIS model on the coauthorship network; $\bm a$ is the minimizer of $\epsilon_2$. 
(e) Double-well system on the scale-free network; $\bm a$ is the leading eigenvector. (f) Double-well system on the scale-free network; $\bm a$ is the minimizer of $\epsilon_2$. (g) Double-well system on the coauthorship network; $\bm a$ is the leading eigenvector. (h) Double-well system on the coauthorship network; $\bm a$ is the minimizer of $\epsilon_2$. 
(i) GLV model on the scale-free network; $\bm a$ is the leading eigenvector. (j) GLV model on the scale-free network; $\bm a$ is the minimizer of $\epsilon_2$. (k) GLV model on the coauthorship network; $\bm a$ is the leading eigenvector. (l) GLV model on the coauthorship network; $\bm a$ is the minimizer of $\epsilon_2$. 
The scale-free network used in (a), (b), (e), (f), (i), and (j) is the same as
the one used in Figs.~\ref{fig:SIS}(a) and \ref{fig:SIS}(b).
The theoretical estimates, i.e., the dashed and dotted lines, for the scale-free network 
shown in (a), (b), (e), (f), (i), and (j) are the same as those shown in 
Figs.~\ref{fig:SIS}(a), \ref{fig:SIS}(b), \ref{fig:double-well}(a), \ref{fig:double-well}(b), \ref{fig:GLV}(a), and \ref{fig:GLV}(b), respectively.}
\label{fig:noisy}
\end{figure}

We show in Figs.~\ref{fig:noisy}(a) and \ref{fig:noisy}(b) the numerical results for the SIS model run on the same scale-free network as the one used in Figs.~\ref{fig:SIS}(a) and \ref{fig:SIS}(b). We verify that the fluctuation in $R$ using the leading eigenvector as $\bm a$, shown by the solid line in Fig.~\ref{fig:noisy}(a), carries less noise than $R$ using the minimizer of $\epsilon_2$ as $\bm a$, shown by the solid line in Fig.~\ref{fig:noisy}(b). We stress that we have obtained Figs.~\ref{fig:SIS}(a) and \ref{fig:SIS}(b) from the same simulation; we have only changed the observable, $R$. The results are qualitatively similar for the coauthorship network, while the difference between the two cases is now smaller (see Figs.~\ref{fig:noisy}(c) and \ref{fig:noisy}(d)). The results for the double-well system, shown in Figs.~\ref{fig:noisy}(e)--(h), and those for the GLV model, shown in Figs.~\ref{fig:noisy}(i)--(l), reinforce our claim that $R$ is more robust against noise when one uses the leading eigenvector rather than the minimizer of $\epsilon_2$ as $\bm a$.

The theoretical estimate based on the spectral method does not depend on the noise (see the dashed and dotted lines in Fig.~\ref{fig:noisy}).
However, the target observable to be approximated, $R$, is noisier when $\bm a$ is a non-leading eigenvector than the leading eigenvector. Therefore, we conclude that the one-dimensional reduction using the leading eigenvector as $\bm a$ better describes the one-dimensional projection of the original network dynamics in the presence of noise unless the magnitude of the noise is small.

\section{Discussion\label{sec:discussion}}

In the present study, we have explored two ideas to try to improve the spectral method to reduce dynamical systems on networks into a one-dimensional dynamics. The first idea is to constrain the use of the Taylor expansion of the dynamical variables, $\{ x_1, \ldots, x_N \}$, around one reference point, $R$. The original spectral method uses the Taylor expansion around $R$ and $\beta^* R$, where $\beta^* \neq 1$ in general. Our Taylor expansion has led to a one-dimensional reduction that does not contain $\beta^*$ (i.e., Eq.~\eqref{eq:modified-reduction-final}).
Our second idea is to use the non-leading eigenvector of the adjacency matrix that minimizes the error. We have obtained explicit expressions of the errors, with which one can systematically search the minimizer of the error, especially that of $\epsilon_2$
(see Eq. ~\eqref{eq:error-modified-reduction}). Note that our method requires the calculations of all the eigenvalues and eigenvectors, which costs $O(N^3)$ time. This is an important limitation of the present method when we apply the method to large networks. In contrast, the previous studies used the leading eigenvector \cite{Laurence2019PhysRevX,Thibeault2020PhysRevResearch}. We have found that, for networks that are relatively homogeneous in the degree, such as the Erd\H{o}s-R\'{e}nyi random graph and the Watts-Strogatz small-world network model,
the leading eigenvector almost always minimizes the errors. In contrast, when the degree is heterogeneously distributed, the optimal eigenvector tended to be a non-leading one, which is particularly the case for larger and sparser networks. For these networks, our modified spectral method is expected to enjoy reduced approximation errors.

We have assessed the performance of approximating the one-dimensional observable, $R$, for three dynamical systems on scale-free networks for which the leading eigenvector and the minimizer of $\epsilon_2$ do not tend to coincide. We have shown that the spectral method using the minimizer of $\epsilon_2$ as $\bm a$ tends to surpass the original spectral method across the different dynamical systems and networks (see Figs.~\ref{fig:SIS}, \ref{fig:double-well}, and \ref{fig:GLV}). However, we have also found that our spectral method using the minimizer of $\epsilon_2$ as $\bm a$ has two essential limitations. First, it is not good at estimating the bifurcation point. In particular, our method locates the bifurcation point extremely far from the correct value if the minimizer of $\epsilon_2$ is associated with an eigenvalue that is much smaller than the largest eigenvalue (see Section~\ref{sub:tradeoff}). In this case, our one-dimensional reduction is qualitatively wrong over a wide range of the bifurcation parameter. Therefore, one cannot accurately approximate $R$ of the original high-dimensional dynamical system in a range of the bifurcation parameter of interest. Second, the spectral method using the minimizer of $\epsilon_2$ as $\bm a$ is not robust against noise. This is because, in the presence of dynamical noise, if one uses a non-leading eigenvector as $\bm a$, the variance of $R$ is larger than in the case of the leading eigenvector used as $\bm a$.
We have also examined the modified spectral method (i.e., $\beta = 1$) with $\bm a$ being replaced by the leading eigenvector. Up to our numerical efforts, this combination is the best performer in that (i) it performs better than the original spectral method (i.e., $\beta = \beta^*$ and the leading eigenvector as $\bm a$) and comparably with our method using the minimizer of $\epsilon_2$ as $\bm a$ in the absence of noise and that (ii) the target $R$ fluctuates less than when $\bm a$ is the minimizer of $\epsilon_2$. The spectral method using $\beta=1$ and the leading eigenvector is also better at locating the bifurcation point than that using the minimizer of $\epsilon_2$. Therefore, we recommend the combination of $\beta=1$ and the leading eigenvector, i.e., Eq. ~\eqref{eq:error-modified-reduction} provided with the leading eigenvector and eigenvalue. We may be able to find pairs of eigenvalue and eigenvector, $(\alpha, \bm a)$, that suppress $\epsilon_2$ much better than the leading eigenvector and the associate eigenvalue is not much smaller than the leading eigenvalue. Using such a pair $(\alpha, \bm a)$ may improve overall performances of the spectral method. Systematically investigating this issue warrants future work.

Both the original and modified spectral methods are based on the Taylor expansion of the differential equations in terms of $x_1$, $\ldots$, $x_N$ around one or multiple common values (e.g., $R$). Therefore, the methods are expected to work better when $\{x_1, \ldots, x_N\}$ is relatively homogeneous. We in fact reached the same conclusion for the GBB reduction in our previous work \cite{KunduKoriMasuda2022PhysRevE}. Our previous numerical simulations suggested that $\{x_1, \ldots, x_N\}$ was not wildly heterogeneous for various dynamical systems including those used in this paper, except near the bifurcation points~\cite{KunduKoriMasuda2022PhysRevE}. However, we do not have a systematic understanding when $\{x_1, \ldots, x_N\}$ is more homogeneous than in other cases. Such an understanding is expected to help us to assess the applicability of the GBB reduction and spectral methods.

The DART is a systematic method to map high-dimensional dynamics on networks into a low-dimensional dynamical system, extending the spectral method \cite{Thibeault2020PhysRevResearch}. The DART is applicable to a diversity of dynamics including synchronization dynamics on networks. Like the original spectral method \cite{Laurence2019PhysRevX}, the DART is based on the Taylor expansion at multiple reference points and the leading eigenvectors of the adjacency matrix. It is interesting to apply the present approach to the DART, in particular to the cases where the reduced dynamics have the dimension larger than one. When the dimension of the reduced dynamics is larger than one, at least one eigenvector with which to take the weighted average of $\{x_1, \ldots, x_N\}$ (e.g., the eigenvector associated with the second largest eigenvalue of the adjacency matrix) will necessarily contain both positive and negative elements due to the Perron-Frobenius theorem and the orthogonality of different eigenvectors. Then, the signal-to-noise ratio for the corresponding observable may be compromised in the presence of dynamical noise (see Section~\ref{sub:noise}). With this possibility being included, it is worth further examining dimension reduction and resilience of noisy dynamical systems on networks.

\section*{Acknowledgments}
N.M. acknowledges support from AFOSR European Office
(under Grant No. FA9550-19-1-7024), the Sumitomo Foundation, the 
Japan Science and Technology Agency (JST) Moonshot R\&D (under Grant No. JPMJMS2021), and the National Science Foundation 
(under Grant No. 2052720).

\appendix

\section{Minimizing the error when the weight vector is an eigenvector of $K$\label{sec:observe-one-variable}}

Let us require Eq.~\eqref{eq:condition-1-pp} to hold exactly instead of Eq.~\eqref{eq:condition-2-pp} in the spectral method.
Then, the weight vector $\bm a$ needs to be an eigenvector of $K$. Because $K$ is a diagonal matrix,
$\bm a$ is a standard unit vector, i.e., $\bm a = \bm e_i \equiv (0, \ldots, 0, \underbrace{1}_{i \text{ th entry}}, 0, \ldots, 0)$ and $\alpha = k_i^{\rm in}$, where $i \in \{1, \ldots, N\}$. This implies that we use a single variable $x_i$ to represent the entire system. Then, it is straightforward to derive $\beta = 1$ and minimize the error for Eq.~\eqref{eq:condition-2-pp} in terms of $\gamma$ as follows:
\begin{equation}
\epsilon_3 \equiv \min_{\gamma} \left\| W^{\top} \bm e_i - k_i^{\rm in} \gamma \bm e_i \right\|^2
= \min_{\gamma} \left[ \sum_{\ell=1}^N (w_{i\ell})^2 + (k_i^{\rm in})^2 \gamma^2 \right]
= \sum_{\ell=1}^N (w_{i\ell})^2,
\end{equation}
which is realized by $\gamma = 0$. We find that $\epsilon_3$ is minimized with respect to $i$ when the $i$th is node has the smallest in-degree in the network. 
The corresponding one-dimensional reduction is given by
\begin{equation}
\frac{\text{d}x_{\overline{i}}}{\text{d}t} = F(x_{\overline{i}}) + k_{\overline{i}}^{\rm in} G(R, 0),
\label{eq:dR/dt-Laurence-with-K-final}
\end{equation}
where $\overline{i}$ is the index of the node with the smallest in-degree.

In the case of our modified spectral method, we require that Eq.~\eqref{eq:new-condition-1-pp} exactly holds instead of
Eq.~\eqref{eq:new-condition-2-pp}. Then, we again obtain $\bm a = \bm e_i$ and $\alpha = k_i^{\rm in}$, where $i \in \{1, \ldots, N\}$.
Then, the error for Eq.~\eqref{eq:new-condition-2-pp} is given by 
\begin{equation}
\epsilon_4 = \left\| W^{\top} \bm a - \alpha \bm a \right\|^2 = \sum_{\ell=1}^N (w_{i \ell})^2 + (k_i^{\rm in})^2.
\end{equation}
If the network is unweighted, we use $(w_{i \ell})^2 = w_{i \ell} \in \{ 0, 1 \}$ to obtain $\epsilon_4 = k_i^{\rm in} + (k_i^{\rm in})^2$, which is minimized when we select $i$ with the smallest in-degree. In this case, the one-dimensional reduction is given by
\begin{equation}
\frac{\text{d}x_{\overline{i}}}{\text{d}t} = F(x_{\overline{i}}) + k_{\overline{i}}^{\rm in} G(R, R).
\label{eq:dR/dt-modified-with-K-final}
\end{equation}

\end{document}